\documentclass{jfm}
\usepackage{xcolor}
\usepackage{graphicx}
\usepackage{epstopdf, epsfig}
\usepackage{url}
\usepackage[colorlinks, allcolors=blue]{hyperref}
\usepackage{soul}
\usepackage{amsmath}

\usepackage{natbib}
\usepackage[colorlinks]{hyperref}
\usepackage{etoolbox}
\usepackage{soul}

\makeatletter

\patchcmd{\NAT@citex}
  {\@citea\NAT@hyper@{%
     \NAT@nmfmt{\NAT@nm}%
     \hyper@natlinkbreak{\NAT@aysep\NAT@spacechar}{\@citeb\@extra@b@citeb}%
     \NAT@date}}
  {\@citea\NAT@nmfmt{\NAT@nm}%
   \NAT@aysep\NAT@spacechar\NAT@hyper@{\NAT@date}}{}{}

\patchcmd{\NAT@citex}
  {\@citea\NAT@hyper@{%
     \NAT@nmfmt{\NAT@nm}%
     \hyper@natlinkbreak{\NAT@spacechar\NAT@@open\if*#1*\else#1\NAT@spacechar\fi}%
       {\@citeb\@extra@b@citeb}%
     \NAT@date}}
  {\@citea\NAT@nmfmt{\NAT@nm}%
   \NAT@spacechar\NAT@@open\if*#1*\else#1\NAT@spacechar\fi\NAT@hyper@{\NAT@date}}
  {}{}

\makeatother

\newcommand{\tabref}[1]{Table ~\ref{#1}}
\newcommand{\figref}[1]{Fig.~\ref{#1}}

\newcommand{\Peclet}[0]{\mathrm{Pe}}
\newcommand{\Nusselt}[0]{\mathrm{Nu}}
\newcommand{\floor}[1]{\left\lfloor #1 \right\rfloor}
\renewcommand{\vec}[1]{\boldsymbol{#1}}
\newcommand{\im}[0]{\mathrm{i}}

\shorttitle{Advection-Diffusion in Potential Flows}
\shortauthor{K. I. McKee \& K. J. Burns}

\title{Steady advection-diffusion in multiply-connected potential flows}

\author{
    Kyle I. McKee\aff{1}, 
    Keaton J. Burns\aff{1,2}
}

\affiliation{
    \aff{1}Department of Mathematics, Massachusetts Institute of Technology, Cambridge, MA 02139, USA
    \aff{2}Center for Computational Astrophysics, Flatiron Institute, New York, NY 10010, USA
}

\begin{document}

\maketitle

\begin{abstract}
We consider the steady heat transfer between a collection of impermeable obstacles immersed in an incompressible 2D potential flow, when each obstacle has a prescribed boundary temperature distribution. 
Inside the fluid, the temperature satisfies a variable-coefficient elliptic partial differential equation (PDE), the solution of which usually requires expensive techniques. 
To solve this problem efficiently, we construct multiply-connected conformal maps under which both the domain and governing equation are greatly simplified. 
In particular, each obstacle is mapped to a horizontal slit and the governing equation becomes a constant-coefficient elliptic PDE. 
We then develop a boundary integral approach in the mapped domain to solve for the temperature field when arbitrary Dirichlet temperature data is specified on the obstacles. 
The inverse conformal map is then used to compute the temperature field in the physical domain. We construct our multiply-connected conformal maps by exploiting the flexible and highly accurate AAA-LS algorithm. 
In multiply-connected domains and domains with non-constant boundary temperature data, we note similarities and key differences in the temperature fields and Nusselt number scalings as compared to the isothermal simply-connected problem analyzed by \cite{choi2005steady}.
In particular, we derive new asymptotic expressions for the Nusselt number in the case of arbitrary non-constant temperature data in singly connected domains at low P\'eclet number, and verify these scalings numerically. 
While our language focuses on the problem of conjugate heat transfer, our methods and findings are equally applicable to the advection-diffusion of any passive scalar in a potential flow.
\end{abstract}

\section{Introduction}

When a passive scalar field $c(\vec{x},t)$, such as temperature or a solute concentration, is transported within a fluid flow $\vec{u}(\vec{x},t)$, it is advected and diffused so that its evolution is governed by 
\begin{equation}\label{eq:unst_addiff}
    \frac{\partial c}{\partial t} + \vec{u} \cdot \nabla c = \alpha \nabla^2 c,
\end{equation}
where $\alpha$ is the diffusivity of $c$. 
Equation (\ref{eq:unst_addiff}) must be supplemented by suitable boundary conditions which depend on the particular problem under consideration. 
If the system has developed into a steady-state, (\ref{eq:unst_addiff}) simplifies to
\begin{equation}\label{eq:st_addiff}
    \vec{u} \cdot \nabla c = \alpha \nabla^2 c. 
\end{equation}
This is simply a linear elliptic equation for $c$, but due to the non-constant nature of the velocity field, obtaining a numerical solution generally requires discretizing the entire fluid. 
To produce solutions with high accuracy, fine meshes are needed and the associated linear algebra problem becomes costly. 
Standard boundary-integral techniques are also difficult to apply directly to (\ref{eq:st_addiff}), since closed forms of the Green's function are not available for generic velocity fields.

In the present paper we restrict our attention to the case where $\vec{u}$ is a steady two-dimensional incompressible potential flow, so that $\vec{u}=\nabla \phi(x,y)$ and $\nabla \cdot \vec{u} = 0$, past a collection of stationary impermeable obstacles, where each obstacle maintains an arbitrary time-independent boundary temperature profile. 
Potential flows serve as accurate models of flow in Hele-Shaw geometries, flows through porous media, and high Reynolds number flows in situations where viscous effects are confined to boundary layers attached to obstacle boundaries. 
Such flows may be represented in the complex plane in terms of a complex potential $W(z)=\phi(z)+\mathrm{i}\psi(z)$, where $W(z)$ is a complex analytic function of $z=x+\mathrm{i}y$ in the fluid and $\psi(z)$ is the harmonic conjugate of $\phi(z)$. 
The complex flow velocity may then be expressed as $u=\overline{dW/dz}$. 

In addition to describing the flow velocity, the complex potential $W(z)$ can be interpreted as a multiply-connected conformal map from the physical domain ($z$-domain) to the \emph{mapped domain} ($w$-domain), also referred to as \emph{streamline coordinates}.
Under this transformation, both the governing equation \eqref{eq:st_addiff} and problem geometry are simplified greatly. 
Each impermeable obstacle is transformed into a horizontal slit, since the streamfunction $\psi$ must be constant on the boundary of such an obstacle, and the fluid region is transformed to the entire complex plane exterior to these slits.
This transformation dates back to \cite{boussinesq1902pouvoir}.
The governing equation \eqref{eq:st_addiff} is transformed into a constant-coefficient equation in the mapped domain which can be studied analytically and with boundary-integral techniques.
Historically, analytical solutions for advection-diffusion have been sought using this approach for potential flows past canonical isolated obstacles such as flat plates, circles, and elliptic cylinders (\cite{boussinesq1905calculation,hsu1965heat}); see  \cite{galante1990applicability} for a general review.

\cite{choi2005steady} studied advection-diffusion past an isothermal simply-connected obstacle of arbitrary shape within a uniform potential flow. 
Using a single-layer boundary integral formulation in the mapped domain, they derived asymptotic expansions in the P\'eclet number for the far-field temperature profile and net heat flux. 
For example, for potential flow past a disk of radius $R$, they found that as the P\'eclet number increases, the Nusselt number scales asymptotically as
\begin{equation}\label{eq:choihighPeNuss}
    \Nusselt \underset{\Peclet \rightarrow \infty}{\sim} 8 \sqrt{\frac{\Peclet}{\pi}}.
\end{equation}
The Nusselt number is defined as the ratio of the total heat flux out of the obstacle to the quantity $\alpha \Delta T$, where $\Delta T$ is the obstacle temperature relative to infinity. 
The P\'eclet number is given by $\Peclet = U R / \alpha$, where $U$ is the far-field (free stream) velocity.
In the limit of small P\'eclet numbers, the study found that
\begin{equation}\label{eq:choloePeNuss}
    \Nusselt  \underset{\Peclet \rightarrow 0}{\sim} -\frac{2\pi}{\gamma+\log{\left(\frac{\Peclet}{4}\right)}},
\end{equation}
where $\gamma$ is the Euler-Mascheroni constant, $0.57721566 \cdots$.
This work complements several fruitful research directions including advection-diffusion limited aggregation (ADLA)\citep{bazant2003dynamics,davidovitch2005average,bazant2005exact,rycroft2016asymmetric} and artificial freezing \citep{alimov1994equilibrium,mukhamadullina1998hysteretic}. 
More recently, \cite{goyette2019microfluidic} studied advection-diffusion in a flow consisting of source/sink singularities, in the absence of a free stream or any flow obstacles.

In the present paper, we consider the more general problem of advection-diffusion in multiply-connected domains.
Additionally, we allow the boundary of each obstacle to possess a spatially varying temperature distribution. 
First, we develop an accurate numerical procedure for computing the potential flow around a generic collection of smooth objects.
This method is enabled by the AAA-LS algorithm \citep{costa2021aaa}, a flexible tool for solving Laplace problems using rational function approximation.
In the mapped domain, we then solve for the temperature field using a mixed single-plus-double-layer boundary integral method that exploits the homogeneous Green's function of the transformed PDE. 

Using these tools, we investigate both similarities and key differences between the behaviour of our solutions and the isothermal, simply-connected problem that was examined by \cite{choi2005steady}.
We first compare the behaviour of the simply-connected variable-temperature problem with the results of \cite{choi2005steady} in the high and low P\'eclet number limits. 
Notably we demonstrate new scalings for the Nusselt number, both numerically and analytically, in the low P\'eclet number limit, based on the structure of the imposed temperature boundary conditions.
We then investigate the high P\'eclet number limit and multiply-connected problems.

The methods developed herein, being highly generic and accurate to many digits of precision, should prove useful for researchers looking to solve advection-diffusion problems in multiply-connected domains. 
Prior work considering nontrivial multiply-connected geometries was highly limited by the lack of flexible and efficient numerical algorithms for these cases.
Our new algorithms for multiply connected conformal maps using AAA-LS, and our efficient boundary integral solvers, are capable of solving many such problems with ease.
Future applications of this line of work might include investigations of multiply-connected ADLA and artificial growth problems, as well as extensions of these techniques to quasi-static advection-diffusion processes.

The remainder of this paper is organized as follows. 
In \S \ref{mathform}, we mathematically describe the multiply-connected conformal map corresponding to the potential flow solution and demonstrate how the governing PDE is simplified under this transformation. 
In \S \ref{numerform}, we formulate our conformal mapping computations as well as the boundary integral technique used to obtain the temperature field in the mapped domain. 
We also derive and verify expressions for the flux from each obstacle. 
In \S \ref{comparison}, we compare our variable temperature results, in both simply and multiply-connected domains, to existing asymptotic results pertaining to the special case of simply-connected isothermal obstacles; in doing so, we demonstrate similarities and key differences in our generalized setting. 
In \S \ref{SCLP}, we derive asymptotic scalings for the Nusselt number, under variable temperature data, in the low P\'eclet number limit. 
In \S \ref{examps}, we showcase the utility of our method by studying advection-diffusion past multiple obstacles over a wide range of P\'eclet numbers. 
We conclude by noting potential future extensions and applications of our results.

\section{Mathematical Formulation}\label{mathform}

Here we formulate the problem of finding the incompressible potential flow comprising a uniform free stream past a collection of impermeable obstacles. 
We then interpret the complex potential of this flow as a conformal mapping from the physical domain to streamline coordinates, and demonstrate how the governing PDE for steady advection-diffusion is greatly simplified in the mapped domain.

\subsection{The Complex Potential, $W(z)$, for Flow Past Multiple Obstacles}\label{comppot}

\begin{figure}
\centering
 \includegraphics[width=\linewidth]{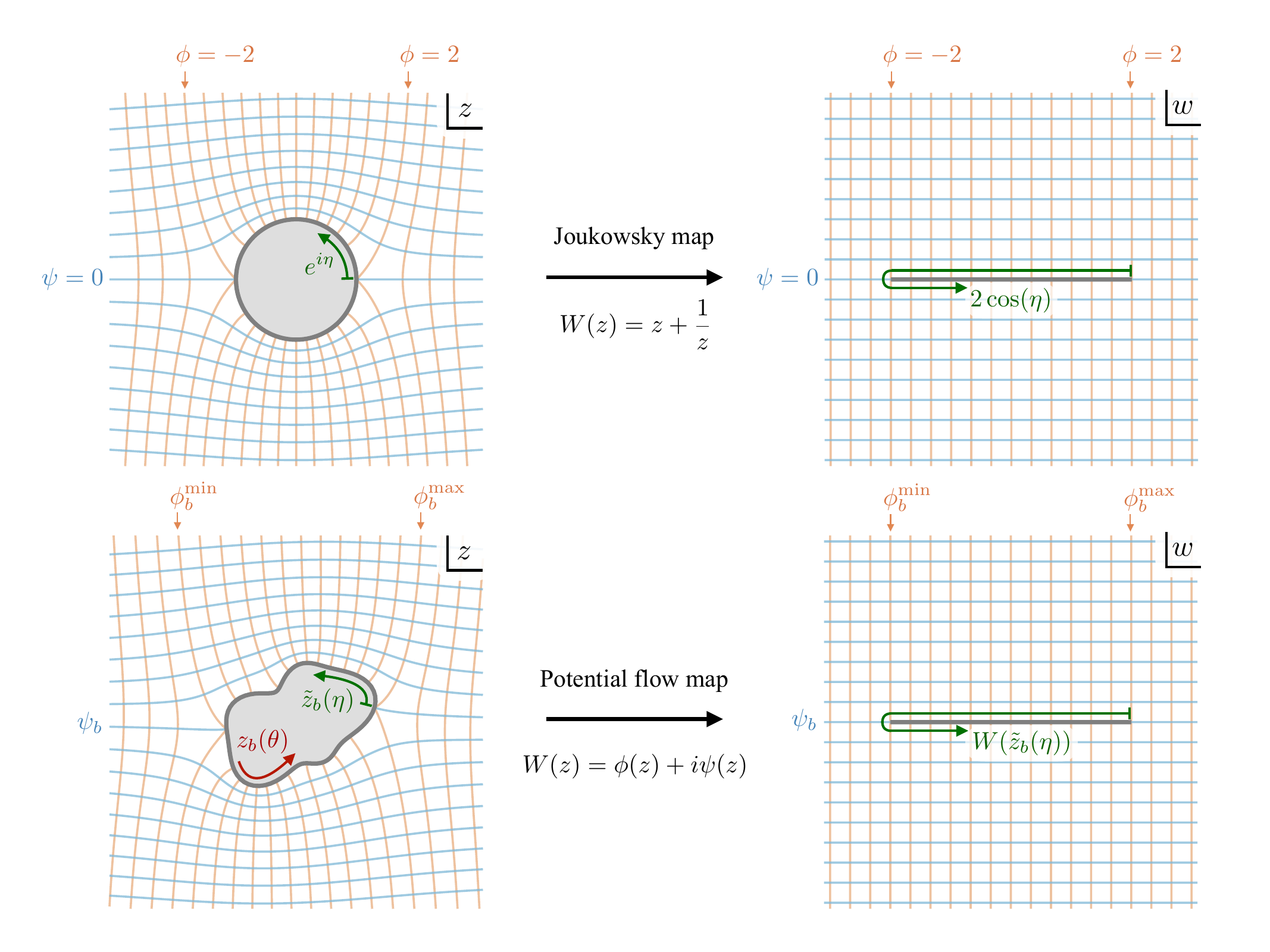}
 \caption{\label{fig:confmapimageplot} Two examples of potential flow solutions around obstacles and their interpretation as a conformal maps to streamfunction coordinates.
 The top panel depicts a free stream flow with $\vec{u}_{\infty} = (1,0)$ around a circle.
 The complex potential is given by the classical Joukowsky map, bringing the unit circle to a slit along the real axis.
 The bottom panel depicts a free stream flow with $\vec{u}_{\infty} = (1,0)$ around a generic smooth obstacle parameterized as $z_j(\theta)$.
 The complex potential is found numerically but also maps the obstacle boundary to a horizontal slit in the mapped domain.
 A parameterization of the obstacle, which we term the Joukowsky parameterization $\tilde{z}(\eta)$, can be found whose image traverses the slit in a fashion like $\cos\eta$, analogous to the case of the circle under the classical Joukowsky map.}
\end{figure}

We now formulate the boundary value problem describing the incompressible potential flow around impermeable obstacles with a prescribed free-stream (far-field) velocity. 
We consider flow past a collection of $N^b$ many obstacles with boundaries given by smooth closed curves $\partial B_j$ for $j \in \{1,\ldots,N^b\}$.
We write the free-stream velocity as $\vec{u}_\infty = (U \cos\Theta, U \sin\Theta)$, where $U>0$ is the free-stream velocity and $\Theta$ is the angle of the stream with respect to the $x$-axis.
We non-dimensionalize lengths by $L$, a characteristic scale of the obstacles, and time by $L/U$.
The non-dimensional potential flow problem is then to find a velocity $\vec{u} = \nabla \phi(x,y)$ such that $\nabla \cdot \vec{u}=0$, $\vec{u} \cdot \vec{n}_j = 0$ where $\vec{n}_j$ is the normal vector along the boundary of each obstacle, and $\vec{u}$ tends to $(\cos\Theta, \sin\Theta)$, with unit magnitude, at infinity.

The incompressibility condition implies $\nabla^2 \phi = 0$, so we proceed by considering the complexified problem with $z = x + \mathrm{i} y$.
The flow potential can then be viewed as the real part of a complex potential $W(z) = \phi(z) + \mathrm{i} \psi(z)$, which is analytic in the fluid $\mathbb{C}\backslash \left(\cup_{k=1}^{N^b}B_k\right)$, and where the streamfunction $\psi(z)$ is the harmonic conjugate of $\phi(z)$.
The complex velocity can then be expressed as $u = \overline{dW/dz}$.
The impermeability conditions can be expressed by requiring the streamfunction to be constant on the boundary of each obstacle.
We thus seek a single-valued complex function $W(z)$ that is analytic in the fluid and has the following properties:
\begin{eqnarray}
\begin{gathered}
    W(z) \sim e^{-\mathrm{i}\Theta}  z,\; |z|\rightarrow \infty \label{eq:farfield} \\
    \mathrm{Im}\left\{W(z)\right\}=\psi_j,\;z\in\partial B_j,\label{eq:comppot}
\end{gathered}
\end{eqnarray}
for some \textit{a priori} unknown constants, $\left\{\psi_1,\ldots, \psi_{N^b} \right\}$. 
The flow is assumed to be free of singularities (e.g.\ source/sinks and point vortices), and further possesses zero-circulation around each obstacle.
Writing $W(z) = e^{-\mathrm{i} \Theta} z + \delta W(z)$, the free-stream component can be subtracted and we can equivalently seek a complex analytic function $\delta W(z)$ satisfying:
\begin{eqnarray}
\begin{gathered}
    \delta W(z) \rightarrow 0, \; |z| \rightarrow \infty \\
    \mathrm{Im}\left\{\delta W(z)\right\} = \psi_j -  \mathrm{Im}\left\{e^{-\mathrm{i} \Theta} z\right\}, \; z \in \partial B_j.\label{eq:delcomppot}
\end{gathered}
\end{eqnarray}

For nontrivial geometries, $W(z)$ typically cannot be attained in a simple closed form.
However, the problem for $\delta W(z)$ can be solved numerically using a least-squares approach over sets of functions that are holomorphic in the fluid and decay at infinity.
For instance, solutions can be found using truncated Laurent series centered within each obstacle.
However, this approach can converge very slowly in some cases, for instance when the boundaries are nonsmooth, slender, or concave.
In these cases, the Laurent series ansatz can be supplemented with additional poles in the obstacle interiors; these poles can greatly enhance the solution accuracy, provided that their locations are chosen appropriately.
We take the approach of using the AAA-LS algorithm \citep{costa2021aaa} to automatically determine efficient pole placements for generic smooth obstacles. 
The numerical details of this approach will be discussed in \S \ref{aaa}.

The complex potential governing the fluid flow, $W(z)$, may alternatively be interpreted as a conformal map between the region exterior to the obstacles in the physical domain and the entire complex plane exterior to a collection of slits, with the slits representing the images of $\partial B_j$ under $W(z)$.
To illustrate the geometry of the mapping, consider the flow past a unit disk with $\Theta=0$, as is depicted in the top of \figref{fig:confmapimageplot}. 
Here the mapping is given by the classical Joukowsky map $W(z) = z + 1/z$, and the image of the fluid region is the entire complex plane exterior to the slit $[-2,2]$ on the real axis, as is depicted on the top right of \figref{fig:confmapimageplot}.
Points on the circle $z=e^{\mathrm{i}\eta}$ are mapped to a double-cover of the slit as $w=2\cos\eta$.
The images of points just outside the circle $z=(1+\epsilon) e^{\mathrm{i}\eta}$ traverse the slit in a positively oriented fashion starting from the trailing edge ($w=2$).

For a generic non-circular obstacle, the interpretation of $W(z)$ as a conformal map is retained.
The obstacle boundary, say given by some $2 \pi$-periodic parameterized curve $z_j(\theta)$, is mapped to a horizontal slit defined by the interval $S_j = [\phi_j^\mathrm{min}, \phi_j^\mathrm{max}] + \im \psi_j$. The length of the slit is $2s_j \equiv \phi_j^\mathrm{max}-\phi_j^\mathrm{min}$, and it is centered at $\phi_j \equiv (\phi_j^\mathrm{max}+\phi_j^\mathrm{min})/2$. 
A new parameterization of the obstacle boundary $\partial B_j$, which we term the \emph{Joukowsky parameterization} $\tilde{z}_j(\eta)$, can be found such that its image under the mapping is $W(\tilde{z}_j(\eta)) = \phi_j + s_j \cos\eta + \im \psi_j \equiv \phi_j(\eta) + \im \psi_j$.
This is illustrated in the bottom half of \figref{fig:confmapimageplot}.
In multiply-connected geometries, the same interpretation is retained, and each obstacle is mapped to a separate horizontal slit, generically of different lengths and at different streamfunction values.
A triply-connected example is illustrated in \figref{fig:threebodies}. 

\subsection{Governing Equation in Streamline Coordinates}\label{eq:streamcoord}

Under the conformal mapping provided by the complex potential, the governing equation for steady advection-diffusion simplifies dramatically.
Non-dimensionalizing lengths by $L$, time by $L/U$, and temperature by $\Delta T$, a characteristic scale for the variation of the obstacle temperatures from the far-field, \eqref{eq:st_addiff} becomes
\begin{equation}\label{eq:nondim_advdiff}
    \Peclet \, \vec{u} \cdot \nabla c = \nabla^2 c,
\end{equation}
where the P\'eclet number $\Peclet=UL/\alpha$ and all quantities are non-dimensional.
This equation can be interpreted as representing the advection and diffusion of $c$ with unit diffusivity by a potential flow with a free-stream velocity of amplitude $\Peclet$, in which case it can be incorporated into the conformal map and hence the obstacle slits in the mapped domain will scale with $\Peclet$.
Alternatively, the equation can be interpreted as representing the advection and diffusion of $c$ with diffusivity $\Peclet^{-1}$ by a potential flow with unit free-stream velocity.
In this case, the conformal map depends only on $\Theta$, and the P\'eclet dependence remains present in the PDE in the mapped domain.
We proceed with the latter approach for computational convenience.

Under the conformal mapping defined by $W(z)$, \eqref{eq:nondim_advdiff} can be rewritten in streamfunction coordinates as
\begin{equation}
    \Peclet \, J \,c_\phi = J(c_{\phi \phi} + c_{\psi \psi})
\end{equation}
where $J = |W'(z)|^2$ is the determinant of the Jacobian of the conformal map.
Only when $\vec{u}$ is a potential flow does the Jacobian factor from the advective term exactly match the Jacobian factor in the Laplacian under this change of coordinates \citep{bazant2004conformal}.
Since this Jacobian is nonzero in the fluid, where $W(z)$ is holomorphic, the Jacobian factors can be canceled and we are simply left with a constant-coefficient equation in the mapped domain:
\begin{equation}\label{eq:advecslitdom}
    \Peclet \, c_{\phi} = c_{\psi\psi}+c_{\phi\phi}.
\end{equation}
The boundary conditions for $c(\phi,\psi)$ are now specified on the boundaries of each obstacle, which are horizontal slits in the mapped domain.

The constant-coefficient nature of \eqref{eq:advecslitdom} dramatically simplifies the PDE's solution, since the free-space Green's function of this operator, and it's normal derivative, can be written explicitly as 
\begin{equation}\label{eq:green}
    G(\phi, \psi) =\frac{1}{2\pi} \exp(\tfrac{\Peclet}{2} \phi) K_0(\tfrac{\Peclet}{2}\sqrt{\phi^2 + \psi^2})
\end{equation}
and
\begin{equation}\label{eq:green_deriv}
    G_\psi(\phi, \psi) = - \frac{\Peclet}{4\pi} \exp(\tfrac{\Peclet}{2} \phi) K_1(\tfrac{\Peclet}{2} \sqrt{\phi^2 + \psi^2}) \frac{\psi}{\sqrt{\phi^2 + \psi^2}},
\end{equation}
where $K_0$ and $K_1$ are modified Bessel functions of the second kind.
Having simple known Green's functions enables the efficient solution of the PDE using boundary integral methods.
Importantly, we note that $G$ is continuous across each slit (at $\psi=0$) while $G_\psi$ is discontinuous.
Thus, to solve problems with arbitrary boundary data, including different Dirichlet values on the top and bottom of each slit, we will adopt a combined single-plus-double-layer boundary integral formulation, where distributions of both (\ref{eq:green}) and (\ref{eq:green_deriv}) are included. 
For completeness, we note that if the P\'eclet scaling is instead incorporated into the conformal map, as described above, the PDE and Green's functions in the mapped domain become independent of P\'eclet.

\subsection{Variable-Temperature Boundary Conditions}

We consider advection-diffusion between obstacles with arbitrary Dirichlet temperature profiles on their boundaries. 
In the physical domain, the temperature conditions are specified on each obstacle boundary $\partial B_j$ as 
\begin{equation} 
    c(z_j(\theta)) = c_j(\theta),
\end{equation}
or, alternately, using the Joukowski parameterizations of each obstacle, as 
\begin{equation} 
    c(\tilde{z}_j(\eta)) = \tilde{c}_j(\eta).
\end{equation}
In the mapped domain, the boundary condition specification is slightly more subtle. 
Each boundary $\partial B_j$ is mapped to a horizontal slit, with the top and bottom of the slit being the images of distinct points on the obstacle boundary in the physical domain. Since the Dirichlet data need not be constant in the physical domain, boundary values on each slit in the mapped domain can thus generically be different when approached from above or below.
These top and bottom boundary conditions can be written simply using the Joukowsky parameterization as
\begin{subeqnarray}
    c(\phi_j(\eta), \psi_j + \epsilon) = \tilde{c}_j(\eta)\, \quad 0 \leq \eta < \pi \\
    c(\phi_j(\eta), \psi_j - \epsilon) = \tilde{c}_j(\eta)\, \quad \pi \leq \eta < 2 \pi
\end{subeqnarray}
for $j \in \{1,\ldots,N^b\}$ and where $\epsilon \rightarrow 0^+$.
Because the temperature is generally discontinuous across each slit in the mapped-domain, it is essential that we include a discontinuous double-layer contribution in our representation of $c(\phi, \psi)$, as will be developed in \S \ref{b_intform}.

\section{Numerical Formulation}\label{numerform}

We have seen that by moving into streamline coordinates, the equations governing conjugate heat transfer in a potential flow can be greatly simplified. 
Namely, the variable coefficient elliptic PDE (\ref{eq:st_addiff}) is transformed into the constant-coefficient PDE (\ref{eq:advecslitdom}).
In simply-connected geometries, conformal maps (and thus potential flow solutions) are analytically known in a variety of geometries (\cite{galante1990applicability}). 
In multiply-connected problems, however, the set of analytically known conformal maps is much more limited. 
In the present section, we describe our numerical formulation and solution of multiply-connected advection-diffusion problems. 
First, we compute the necessary conformal transformations by leveraging the AAA algorithm to efficiently compute the complex potential, $W(z)$, for flows around arbitrary numbers of obstacles.
We then formulate a boundary integral solution to the advection-diffusion problem in the mapped domain. 
Finally, the solution is mapped backed to the physical domain via the inverse conformal mapping to $W(z)$.

\subsection{Computation of Conformal Maps}

The class of potential flow geometries for which the complex potential, $W(z)$, can be found in a simple analytical closed form is somewhat limited. 
In simply-connected geometries, there exist a wide variety of known conformal maps from the unit circle to geometries of interest, such as ellipses and polygons \citep{nehari2012conformal}, which thus allow for a closed-form representation of $W(z)$ for those domains. 
However, in multiply-connected geometries, the class of known conformal maps remains limited, despite recent developments that have unlocked a variety of mappings using the Prime function framework of \cite{crowdy2020solving}. 
To address this issue, here we develop a general numerical method for deducing $W(z)$ for an arbitrary collection of $N^b$ obstacles via series solutions supplemented by specifically placed simple poles.
We restrict our consideration to smooth obstacles that are well-separated.
In this regime, the potential flow can be efficiently solved to high precision using the AAA-LS algorithm as is described below. 
However, this method does not expedite convergence rates in close-contact problems. 
Extensions of the AAA-LS method to handle such problems is an open research problem and we do not attempt to address this issue in the present work. 
We thus restrict our attention to well-separated obstacles, specifically ones that are separated by at least their local radii of curvature. 

\subsubsection{Laurent Series Representation}\label{aaa}

The most straightforward representation of $\delta W(z)$ is as a collection of truncated Laurent series centered within each obstacle and containing only negative powers to ensure that $\delta W(z)$ is analytic in the fluid and decays at infinity. 
The form of $W(z)$ is thus given by
\begin{equation}\label{eq:W_ansatz}
    W(z)\approx  \, e^{-\mathrm{i}\Theta}z+\sum_{j=1}^{N^b}\sum_{k=1}^{N^L_j}\frac{a_{j,k}}{\left(z-z_j\right)^k},
\end{equation}
where $z_j$ is a point centered within $B_j$ and $N^L_j$ is the degree of truncation of the Laurent expansion centered within the $j^{\mathrm{th}}$ obstacle. 
This representation can be used to find an approximation of $W(z)$ by solving a least-squares problem for the series coefficients $\{a_{j,k}\}$ by enforcing the boundary conditions $\mathrm{Im}\{W(z)\}=\psi_j$ at $N^s_j$-many sample points $z^s_{j,k}$ along each obstacle boundary $\partial B_j$. In practice, this least-squares solve converges best when $N^s_j \sim 6 N^L_j$ and the system is solved using Arnoldi orthogonalization \citep{brubeck2021vandermonde}. 

In theory, when the full Laurent series is retained, and $N_j^L$ tends to infinity, the equation holds exactly for obstacles without corners.
Unfortunately, when the obstacles have regions of high radius of curvature or include concavities, the Laurent series representation can converge extremely slowly \citep{trefethen2020numerical,trefethen2024polynomial}. 
For instance, to attain a mapping $W(z)$ to an accuracy of $10^{-12}$ for the irregular shape in \figref{fig:confmapimageplot}, the number of Laurent series coefficients that must be retained grows considerably to $N_j^L=340$.
Such large $N_j^L$ invariably makes solving for the series coefficients, as well as subsequent evaluations of $W(z)$, slow, so alternate strategies are needed for all but the simplest geometries.

\subsubsection{Supplementing Laurent Series with Simple Poles}

For nonsmooth or nonconvex obstacles, supplementing the Laurent series with a set of strategically placed poles -- a form of the method of fundamental solutions (MFS) -- substantially improves the convergence of rational approximation of exterior harmonic functions.
The key challenge is in finding a set of suitable pole locations.
When obstacles have sharp corners, \citet{gopal2019solving} showed that an exponentially clustered set of poles near corners resolves corner singularities with root-exponential convergence; this result is deeply connected to the work of \cite{newman1964rational}. 
When the boundary shape is smooth (i.e., it does not contain sharp corners), a set of suitable poles can still be attained by exploiting the AAA-LS algorithm \citep{costa2021aaa}. 
The AAA algorithm computes a rational barycentric approximation of a function defined on a contour in the complex plane \citep{nakatsukasa2018aaa}.
The poles of this rational function depend on the function being approximated as well as the contour on which the function is sampled. 
The AAA-LS algorithm computes the AAA approximation to some boundary data on $\partial B_j$ and uses the interior poles of the resulting rational function as MFS support points.
Here we apply this algorithm to the Schwarz function, $S(z) = \overline{z}$ for $z \in \partial B_j$, on obstacle boundaries (for more on approximation of the Schwarz function, see \citep{trefethen2025numerical}).
In practice, this method seems to automatically produce efficient pole placement locations for solving Laplace problems in many geometries.

Therefore, to efficiently produce multiply-connected potential flow solutions, we supplement low-order Laurent series within each obstacle with simple poles produced via the AAA-LS procedure.
We determine the poles by applying AAA to the Schwarz function of each obstacle independently and retaining the $N_j^p$-many interior poles; we denote the collection of all such poles by $\{z^p_k\}$ with $k=1,\ldots,N^p$. 
Such poles are depicted in \figref{fig:potato} and \figref{fig:threebodies}.

The new AAA-LS representation of $W(z)$ then takes the form
\begin{equation}\label{eq:W_suppoles}
    W(z) \approx  e^{-\mathrm{i}\Theta}z+\sum_{j=1}^{N^b}\sum_{k=1}^{N^L_j}\frac{a_{j,k}}{\left(z-z_j\right)^k}+\sum_{k=1}^{N^p}\frac{b_k}{z-z^p_k},
\end{equation}
where now the coefficients $\{a_{j,k}\}$ and $\{b_k\}$ are simultaneously found through the solution of the associated least-squares problem.
For example, to compute a mapping $W(z)$ to an accuracy of $10^{-12}$ for the irregular shape in \figref{fig:confmapimageplot}, only $N^L = 20$ Laurent series terms are needed in addition to the $N^p=42$ simple poles with locations furnished by AAA.
Without the AAA poles, as was mentioned earlier, $N^L=340$ Laurent series terms must be retained to achieve the same accuracy.

While the poles found via AAA efficiently represent data on each independent non-circular boundary, we have found that AAA poles do not help to resolve close-contact problems.
Additionally, solving the AAA problem globally tends to produce lines of poles in the fluid region between nearby obstacles, but these poles are discarded and therefore still can not help in the least-squares step.
Future modifications of this approach, for instance incorporating points from alternate fundamental-solution techniques (e.g.\ \cite{cheng_method_1998}), or placing higher-order poles at the AAA-LS simple pole locations, may extend its capabilities for close-contact problems, but we leave these explorations to future work.

\subsubsection{Solving the Least Squares Problem}

To solve for the coefficients $\{a_{j,k}\}$ and $\{b_k\}$ in the representation of $W(z)$ defined in (\ref{eq:W_suppoles}), we substitute (\ref{eq:W_suppoles}) into (\ref{eq:comppot}).
Sampling the equation at points along the domain boundary defines a linear least squares problem for the coefficients. 
We denote the boundary sample points on the $j^{\mathrm{th}}$ obstacle boundary $\partial B_j$ as $\{z^s_{j,k}\}$ for $k\in\{1,\ldots,N^s_j\}$, where $N^s_j$ is the total number of sample points on $\partial B_j$.
The constraint associated with the $k^{\mathrm{th}}$ sample point on the $j^{\mathrm{th}}$ obstacle may be expressed simply as 
\begin{equation}
    \mathrm{Im}\left\{W(z^s_{j,k})\right\}\approx \psi_j.
\end{equation}
These equations are imposed for each sample point on each obstacle, forming a linear least squares problem for the representation coefficients, $\{a_{j,k}\}$ and $\{b_k\}$, along with the a priori unknown streamfunction values $\{\psi_j\}$.
Performing Arnoldi orthogonalization on the Laurent series portion of the matrix prevents ill-conditioning of the associated Vandermonde blocks \citep{brubeck2021vandermonde}. 
For more details of the associated linear algebra problem, see \cite{baddoo2020lightning}. 
Therein, \cite{baddoo2020lightning} places poles manually whereas we use the AAA algorithm to determine the pole locations; however, the linear algebra problem remains the same.
With the proper Arnoldi orthogonalization, the system can be solved using standard linear least squares techniques.
The error in the resulting map can be assessed simply by evaluating the residual $|\mathrm{Im}\{W(z)\}-\psi_j|$ at additional points along each boundary $z \in \partial B_j$, and $N_j^L$ can be increased until the convergence is suitable.
In practice, we find using $N^s_j \sim 6 (N^L_j + N^p_j)$ works well.

\subsection{Joukowsky parametrization}\label{jouk}

The numerical conformal map $W(z)$ solved above maps each obstacle $\partial B_j$ in the $z$-domain to a horizontal slit $S_j$ in the $w$-domain, where the PDE is constant-coefficient and can be efficiently solved using boundary integral techniques.
As a first step in solving the boundary integral equation in the $w$-domain, it is useful to reparametrize the obstacles after computing the flow map $W(z)$.
The obstacles are initially prescribed with some positively-oriented shape and temperature parameterizations as $z_j(\theta)$ and $c_j(\theta)$, which are generic parametrizations with no restrictions, other than that they are smooth $2\pi$-periodic functions of $\theta$.

The flow map $W(z)$ maps the boundary of each object to a horizontal slit in the $w$-domain described by $S_j = \{\phi + \im \psi_j : \phi^{\mathrm{min}}_j \leq \phi \leq \phi^{\mathrm{max}}_j\}$.
Motivated by the Joukowsky map of a circle to the slit, we seek a reparametrization $\theta_j(\eta)$ for each obstacle such that 
\begin{equation}
    W(z_j(\theta_j(\eta))) = \underbrace{\underbrace{\frac{\phi_j^\mathrm{min}+\phi_j^\mathrm{max}}{2}}_{\phi_j} +\underbrace{\frac{\phi_j^\mathrm{max}-\phi_j^\mathrm{min}}{2}}_{s_j} \cos(\eta)}_{\phi_j(\eta)} + \im \psi_j.
\end{equation}
This takes the form of the image of the Joukowsky map of the circle parametrized as $e^{i\eta}$, stretched by $s_j$ and shifted by $\phi_j(0) + \im \psi_j$.
Importantly, it provides a parametrization for the obstacle in the $z$-domain that conforms to the values of $\phi$ on the boundary under the conformal map.

The parametrization is unique, and maintains the orientation of the original parametrization, if we specify 
\begin{equation}
    \theta_j(\eta) = \eta + \delta \theta_j(\eta)
\end{equation}
where $\delta \theta_j(\eta)$ is a $2\pi$-periodic function of $\eta$, and take $\theta_j(0) = \mathrm{argmax}_{\theta \in [0, 2\pi)} \mathrm{Re}\left\{ W(z_j(\theta))\right\}$.
A high-precision Fourier approximation of $\delta \theta_j(\eta)$ is found using ApproxFun.jl and rootfinding to determine $\delta \theta_j(\eta)$ for various $\eta$.
We use this function to compute the Joukowsky parametrizations, $\tilde{z}_j(\eta) = z_j(\theta_j(\eta))$ and $\tilde{c}_j(\eta) = c_j(\theta_j(\eta))$, for each obstacle.
Numerically, these are each represented with high-precision Fourier approximations using ApproxFun.jl.

\subsection{Boundary integral formulation}\label{b_intform}

With the flow-induced Joukowsky parametrizations at hand, we formulate the boundary integral representation for the temperature in the $w$-domain using a mixed single/double-layer representation:
\begin{equation}\label{eq:T_rep}
    c(\phi, \psi) = \sum_{j=1}^{N^b} S[f_j,s_j](\phi-\phi_j,\psi-\psi_j) + D[g_j,s_j](\phi-\phi_j,\psi-\psi_j),
\end{equation}
where the single/double-layer operators are given by
\begin{equation}
    S[f,s](\phi,\psi) = \int_{-s}^s f(s') G(\phi-s', \psi) ds',
\end{equation}
\begin{equation}
    D[g,s](\phi,\psi) = \int_{-s}^s g(s') G_\psi(\phi-s', \psi) ds',
\end{equation}
\noindent and the Green's function and its vertical/normal derivative are given by \eqref{eq:green} and \eqref{eq:green_deriv}.
We note that the single-layer operator $S$ is continuous across $\psi=0$, while the double-layer operator $D$ has limits $D[g,s](\phi, 0^\pm) = \mp \tfrac{1}{2} g(\phi)$ for $|\phi| < s$.
We include both terms in our representation so that generic variable temperature conditions, corresponding to different limits approaching each slit from above and below, can be applied on the obstacles.

The single and double layer densities, $f_j$ and $g_j$, are determined by requiring the temperature to approach the prescribed temperature boundary conditions $c_j$ on the top and bottom of each slit:
\begin{gather}
    \lim_{\epsilon \rightarrow 0^+} c(\phi_j + s_j \cos(\eta), \psi_j + \epsilon) = \tilde{c}_j(\eta), \quad 0 \leq \eta < \pi \\
    \lim_{\epsilon \rightarrow 0^+} c(\phi_j + s_j \cos(\eta), \psi_j - \epsilon) = \tilde{c}_j(\eta), \quad -\pi \leq \eta < 0
\end{gather}
Subtracting these equations provides local equations determining the double-layer density on each slit, since the single layer contributions (and double-layer contributions from other slits) are continuous across each slit:
\begin{subeqnarray}\label{eq.DL_solve}
    \tilde{c}_j(\eta) - \tilde{c}_j(-\eta) &=& \lim_{\epsilon \rightarrow 0^+} c(\phi_j(\eta), \psi_j + \epsilon) -  c(\phi_j(\eta), \psi_j - \epsilon) \\
    &=& D[g_j,s_j](s_j \cos(\eta), 0^+) -  D[g_j,s_j](s_j \cos(\eta), 0^-) \\
    &=& - g_j(s_j \cos(\eta))
\end{subeqnarray}
That is, the double-layer densities $g_j$ can be simply and individually determined by the odd part of $\tilde{c}_j(\eta)$ for each obstacle, corresponding to the sine part of the function's Fourier series.
Adding the top and bottom conditions on each slit then produces coupled boundary integral equations for the single layer densities:
\begin{subeqnarray}\label{eq.SL_solve}
    \tilde{c}_j(\eta) + \tilde{c}_j(-\eta) &=& \lim_{\epsilon \rightarrow 0^+} c(\phi_j(\eta), \psi_j + \epsilon) + c(\phi_j(\eta), \psi_j - \epsilon) \\
    &=& 2 \sum_{j'} S[f_{j'},s_{j'}](\phi_j(\eta)-\phi_{j'}, \psi_j-\psi_{j'}) \\ &+& 2 \sum_{j' \neq j} D[g_{j'},s_{j'}](\phi_j(\eta)-\phi_{j'}, \psi_j-\psi_{j'})
\end{subeqnarray}
These are $N^b$-many coupled first-kind Fredholm integral equations for $f_j$, where the inhomogeneous term consists of the even/cosine parts of $\tilde{c}_j(\eta)$ minus the double-layer contributions from the other slits.

\subsection{Boundary integral discretization}

From its identification with the sine part of $\tilde{c}_j(\eta)$, the double-layer densities can be written as $g_j(s) = \tilde{g}_j(s) \sqrt{s_j^2 - s^2}$, where $\tilde{g}_j(s)$ is a smooth function on the interval $[-s_j,s_j]$.
From the theory of first-kind log kernel integral equations on open arcs \citep{atkinson1991}, the single-layer densities can be written as $f_j(s) = \tilde{f}_j(s) / \sqrt{s_j^2 - s^2}$ where $\tilde{f}_j(s)$ is a smooth function on the interval $[-s_j,s_j]$.
We therefore numerically represent each $\tilde{f}_j$ and $\tilde{g}_j$ as Chebyshev series over $[-s_j, s_j]$, and solve for them (via \eqref{eq.DL_solve} and \eqref{eq.SL_solve}, respectively) using Chebyshev collocation.
The single-layer integrals are discretized using a Lagrange polynomial expansion for $f_j$ over the Chebyshev nodes, and the integrals are evaluated to high precision using QuadGK.jl.
The resulting system is solved to determine the Chebyshev representation of $f_j$.

Once $g_j$ and $f_j$ have been determined, the temperature at any point in the $w$-domain can be computed using \eqref{eq:T_rep}.
The integrals are again efficiently evaluated using QuakGK.jl, which adaptively splits the integrand to resolve the singularities introduced by the kernel.
The temperature at any point in the $z$-domain can be computed simply by applying the conformal map $W(z)$ and then evaluating $c$ at the corresponding position in the $w$-domain.

\begin{figure}
    \centering
    \includegraphics[width=\linewidth]{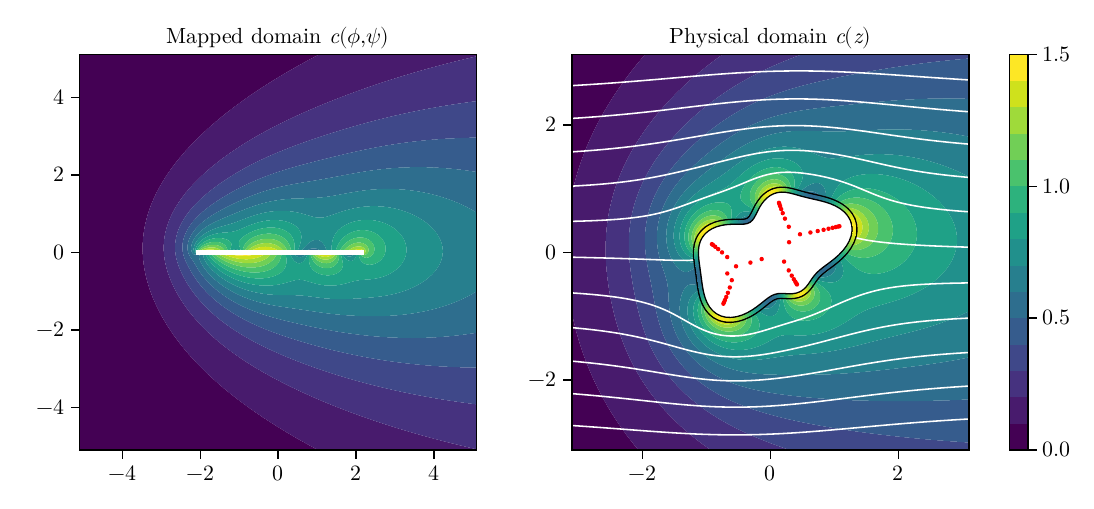}
    \caption{Steady-state temperature under advection-diffusion by a potential flow with $\Peclet=1$ past a single obstacle with a varying prescribed surface temperature. 
    Left: solution in the $w$-domain, where the obstacle takes the form of a horizontal slit. 
    A boundary-integral representation in this domain is used to solve the PDE. 
    Right: solution in the physical $z$-domain, produced by composing the conformal map with the boundary integral solution in the $w$-domain. 
    White lines indicate the streamlines of the potential flow, and the colored line on the border of the obstacle indicates the prescribed temperature boundary condition. 
    The red dots are the poles used to solve for the potential flow via the AAA-LS algorithm.}
    \label{fig:potato}
\end{figure}

\begin{figure}
    \centering
    \includegraphics[width=\linewidth]{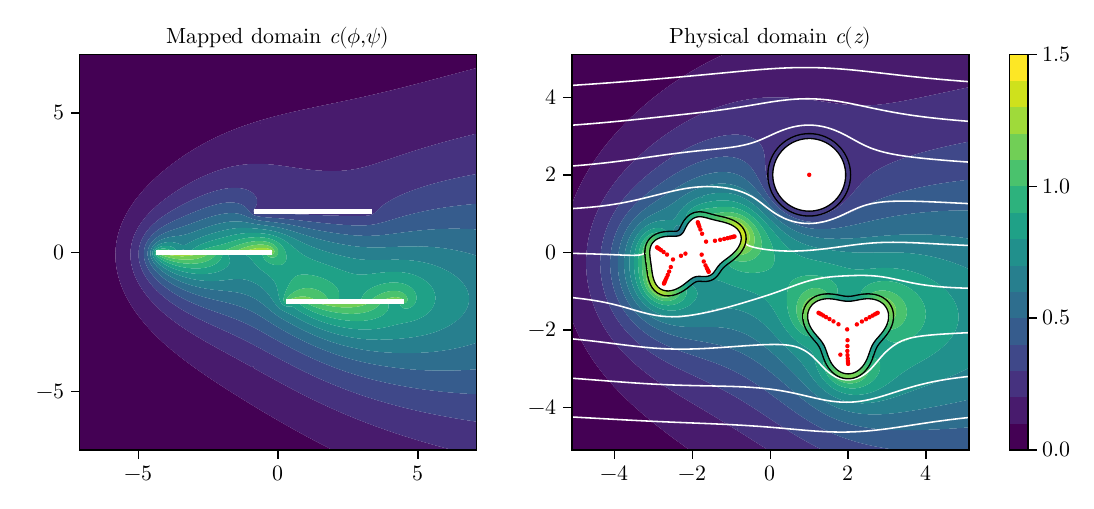}
    \caption{Steady-state temperature for under advection-diffusion by a potential flow with $\Peclet=1$ past three obstacles with various prescribed surface temperatures.
    Left: solution in the $w$-domain, where each obstacle takes the form of a horizontal slit. 
    A boundary-integral representation in this domain is used to solve the PDE. 
    Right: solution in the physical $z$-domain, produced by composing the conformal map with the boundary integral solution in the $w$-domain. 
    White lines indicate the streamlines of the potential flow, and the colored lines on the borders of the obstacles indicate the prescribed temperature boundary conditions. 
    The red dots are the poles used to solve for the potential flow via the AAA-LS algorithm.}
    \label{fig:threebodies}
\end{figure}

We demonstrate this solution procedure in \figref{fig:potato}, where we solve for advection-diffusion around a single obstacle with a varying surface temperature.
The figure shows the solution in both the $w$-domain and the $z$-domain.
\figref{fig:threebodies} shows the solution for potential flow around three obstacles, again in both the $w$ and $z$ domains.

\subsection{Nusselt number computation}
The net flux out of a given obstacle in the physical domain can be computed by integrating the normal component of the flux density, $\boldsymbol{q}=-\alpha\nabla c + c \boldsymbol{u}$, around the boundary. 
Note the advective contribution is zero here since we are taking impermeable boundary conditions ($\vec{u} \cdot \vec{n} = 0$) on each obstacle.
Written in the physical domain, the flux out of the $j^{\mathrm{th}}$ obstacle is given by
\begin{equation}\label{eq:fluxphys}
    \mathcal{F}_j = \int_{\partial B_j} \vec{q} \cdot \vec{n} \,dl = - \alpha \int_{\partial B_j} \frac{\partial c}{\partial n} \,dl,
\end{equation}
where $\vec{n}$ is the unit normal vector and $n$ is the normal distance to the boundary, so that $\partial c/\partial n=\nabla c\cdot \boldsymbol{n}$. 
In streamline coordinates, because each boundary corresponds to a surface of constant $\psi$, we have for the top of the slit that $\left(\partial c / \partial n\right) = |\partial \psi /\partial n|(\partial c/\partial \psi)=|J|^{\frac{1}{2}}(\partial c/\partial \psi)$, whereas for the bottom of the slit we have $\left(\partial c / \partial n\right) =-|J|^{\frac{1}{2}}(\partial c/\partial \psi)$. 
The distance along the boundary in the mapped domain transforms as $dl=|J|^{-\frac{1}{2}}d\phi$.
The Nusselt number associated with the $j^{\mathrm{th}}$ obstacle can then be written in terms of contributions from the top and bottom of the slit in the $w$-domain as
\begin{equation}\label{eq:fluxmapped}
    \Nusselt_j= - \int_{\phi_j^\mathrm{min}}^{\phi_j^\mathrm{max}}\left(\frac{\partial c}{\partial \psi}\Big|_{\psi=\psi_j+\epsilon}-\frac{\partial c}{\partial \psi}\Big|_{\psi=\psi_j-\epsilon}\right)d\phi.
\end{equation}
Using the representation in \eqref{eq:T_rep}, the double-layer contributions to the flux cancel and \eqref{eq:fluxmapped} further simplifies to
\begin{equation}\label{eq:flux_form}
    \Nusselt_j = \int_{\phi_j^\mathrm{min}}^{\phi_j^\mathrm{max}} f_j(\phi-\phi_j) d\phi = \int_{-s_j}^{s_j} f_j(s) ds.
\end{equation}
To verify the validity of (\ref{eq:flux_form}), we consider the test case of two circles of equal radii, $R$, centered on the coordinates $z_j = (-1)^j \im d$ and with temperatures $c_j= (-1)^j \Delta T/2$ for $j=\{1,2\}$. 
In the absence of fluid flow, the net heat flux between the circles is given by $\mathcal{F}=2\pi\Delta T/(\log{\left(1/\rho\right)})$, where $\rho=(d-\sqrt{d^2-R^2})^2/R^2$. 
Thus, in the limit of $\Peclet \rightarrow 0$, the flux in and out of the top and bottom circles, respectively should approach the conductive heat flux.
We solve this problem numerically and find precise agreement with relative flux errors $\left(|Nu_1|-\mathcal{F}\right)\approx 0.7 \, \Peclet$ for $10^{-6} \leq \Peclet \leq 10^{-2}$.

\section{Results}\label{comparison}

Here we utilize the combined conformal remapping and boundary integral solver to produce solutions to a variety of variable-temperature and multiply-connected conjugate heat transfer problems, where the primary result of interest is the heat flux in or out of each obstacle under different flow conditions.
First we examine the heat flux out of simply-connected regions with variable temperature data at both large and small P\'eclet numbers.
We then examine several multiply-connected problems.

\subsection{Joukowsky-Fourier Decomposition of Variable Temperature Data}

To determine the heat flux from a single obstacle with variable temperature data, we first map the problem onto the canonical domain flow around a disk.
For a single generic obstacle $B_1$ in a uniform potential flow, there is a unique conformal map from the fluid region to the exterior of a disk of that maps infinity to infinity with unit derivative.
The radius of the resulting disk $R$ is the \emph{conformal radius} of $B_1$.
Since this map is unique, it must be identical to the map formed by composing the conformal map $W(z)$ with the inverse of the standard Joukowsky map stretched by $R$.
The original heat transfer problem therefore corresponds identically to that of a disk of radius $R$ with boundary data given by the Joukowsky parametrization of the original boundary data, i.e.\ $\tilde{c}_1(\eta)$.
We can therefore study the properties of any simply connected variable-temperature problem via its decomposition into Fourier modes on the disk given by the Joukowsky reparametrization of the surface temperature.
We can further restrict our study to Fourier modes on the unit disk, noting that the conformal radius $R$ can be absorbed into a rescaling of the P\'eclet number when $R \neq 1$.

\begin{figure}
    \centering
    \includegraphics[width=\linewidth]{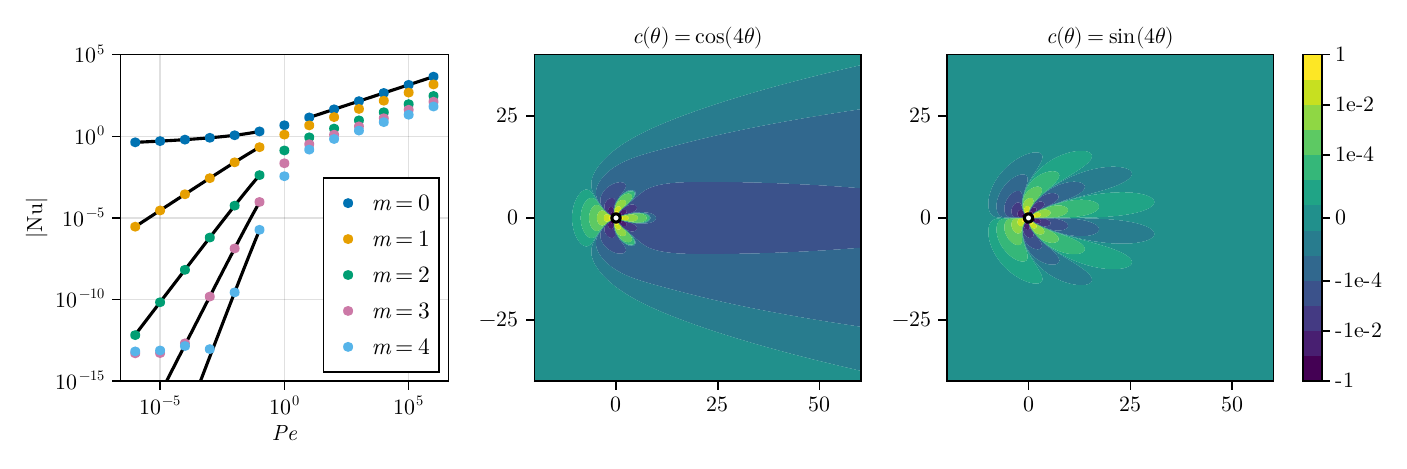}
    \caption{Left: Heat flux out of a unit disk with boundary temperature $\cos{\left(m \theta\right)}$ at various $\Peclet$ numbers, computed using our numerical boundary integral framework (coloured circles).
    The black line on the right indicates the high $\Peclet$-number flux scaling $\sim\Peclet^{1/2}$ derived by \cite{choi2005steady} for isothermal obstacles, \eqref{eq:choihighPeNuss}. 
    Our low-$\Peclet$ asymptotic predictions for the flux given by \eqref{eq:FluxM} – and written explicitly in \tabref{tab:coefftable}– are plotted as solid black lines.
    The numerical results follow these scalings well until numerical errors appear past the adaptive quadrature precision level of $10^{-12}$.
    In the center and right panels, we plot the temperature field induced by boundary temperatures $\cos{\left(4\theta\right)}$ and $\sin{\left(4\theta\right)}$ with $\Peclet = 1$.}
    \label{fig:lowpecosinesplot}
\end{figure}

We therefore consider the canonical case of potential flow with $\Theta=0$ around a unit disk with boundary $z_1(\theta) = e^{\im \theta}$ and boundary temperature given by $c_1(\theta)= \sum_{m=0}^\infty a_m \cos(m\theta) + \sum_{m=1}^\infty b_m \sin(m\theta)$.
Since the heat transfer problem is linear in the Dirichlet data, we can consider each Fourier component independently.
For each sine mode, the temperature data is odd under reflections across the real axis, meaning the temperature values along the top and bottom of the slit in the mapped domain are equal and opposite.
Therefore the single layer density $f_1=0$ for all sine components, and these cases contribute no net heat flux out of the obstacle.
The cosine components are therefore the only part of the generic temperature distribution that contribute to the heat flux.
In \figref{fig:lowpecosinesplot}, we plot the heat flux out of the unit disk as a function of the P\'eclet number for several cosine temperature distributions.
Power law scalings are apparent in both the low and high P\'eclet limits, and we will explore these further in the following sections.
The temperature distributions for cosine and sine distributions with $m=4$ are also shown, and it visually apparent that the antisymmetric cancellation in the sine case prevents net heat transfer to infinity by the potential flow.

\subsection{High P\'eclet Number Flux from Single Obstacles}\label{SCHP}

\cite{choi2005steady} find that the flux from a single isothermal obstacle scales as $\Peclet^{\frac{1}{2}}$, in the limit of $\Peclet \rightarrow \infty$. 
In agreement with this result, we have shown numerically in \figref{fig:lowpecosinesplot} that the flux from a circle with a cosinusoidal boundary temperature also follows the same scaling. 
For boundary data of the form $\cos{\left(m\theta\right)}$, we observe a scaling of $\Nusselt \sim C_m \Peclet^{\frac{1}{2}}$ as $\Peclet\rightarrow \infty$; we have not analytically determined the constants, $C_m$.
We therefore expect this scaling to apply to generic temperature distributions, whose flux can be determined from the distribution's cosine series by linearity, other than the exceptional cases where this leading order contribution cancels. 
Specifically, boundary data of the form $\sum_{k=1}^{M}\alpha_k \cos{\left(k\theta\right)}$ should lead to a Nusselt number scaling of $\Peclet^{\frac{1}{2}}\sum_{k=1}^{M}\alpha_k C_k$. 
Thus it is possible to cancel the leading order behaviour through an appropriate choice of coefficients $\alpha_k$, in which case we expect the flux to scale like $\Peclet^{-\frac{1}{2}}$ as $\Peclet\rightarrow \infty$. 
We note that further analytical progress may be possible to compute the exact limiting flux for each cosine mode, but we leave such calculations to future work.

\subsection{Low P\'eclet Number Flux from Single Obstacles}\label{SCLP}

In the case of low P\'eclet number flows, there are significant differences between variable temperature problems and the isothermal problem analyzed by \cite{choi2005steady}. In the Laplace problem exterior to a unit circle, $c(\infty)$ is simply the average of the Dirichlet data on the circle boundary. 
It follows that any boundary data on the circle possessing a non-zero average cannot satisfy the boundary condition $c(\infty)=0$. 
Therefore the constant boundary data problem examined by \cite{choi2005steady} has the property that the limit of $\Peclet\rightarrow 0$ is singular: there is no Laplace solution in the case of $\Peclet=0$ exactly.
For our canonical cases with boundary data $\cos(m\theta)$ on the unit disk with $m > 0$, a Laplace solution does exist when $\Peclet=0$ and the limit $\Peclet\rightarrow 0$ is not singular. 
We thus deduce that the heat flux in these cases must decay at least as fast as $\it{O}\left(\Peclet\right)$ as $\Peclet\rightarrow 0$. 
This rate is very fast compared to the slow logarithmic decay of the flux for isothermal obstacles (\ref{eq:choloePeNuss}). 
In subsequent sections, we will confirm this prediction by deriving a multipole expansion for the Nusselt number in the $\Peclet\rightarrow 0$ limit.

\subsubsection{Boundary integral analysis}\label{perturb}

Recall that the Joukowsky transformation, $W(z) \equiv \phi+\mathrm{i}\psi=\left(z+\tfrac{1}{z}\right)$, takes the circle $e^{i \theta}$ in the physical domain to the horizontal slit defined by $\phi \in \left[-2,2\right]$ and $\psi=0$, where $\phi=2\cos(\theta)$.
Noting the identity $\cos{\left(m \theta\right)}=T_m(\cos{\left(\theta\right)})$, where $T_m(x)$ is the Chebyshev polynomial of the first kind of degree $m$, the boundary condition $\cos{\left(m \theta\right)}$ on the circle is translated to $T_m(\phi/2)$ on the slit. 
Our goal is now to deduce the flux associated with each such mode. 
In the slit domain, the single-layer equation is
\begin{equation}\label{eq:templowpeinteq}
    \frac{1}{2\pi}\int_{-2}^2 e^{\frac{\Peclet}{2}(\phi-s')}K_0 \left(\frac{\Peclet}{2}|\phi-s'|\right)  f(s') ds' = T_m\left(\frac{\phi}{2}\right).
\end{equation}
For convenience in computing forthcoming integrals, we rescale the slit to the interval $(-1,1)$ by scaling the coordinates such that $x = \phi/2$ and defining $\sigma(s ) = 2 f(2s)$. 
The integral equation then becomes,
\begin{equation}\label{eq:lowpeinteq}
    \int_{-1}^1 e^{\Peclet(x-s')}K_0(\Peclet|x-s|)  \sigma(s) d s\equiv \int_{-1}^1 \tilde{G}(x-s)  \sigma(s) d s = 2\pi T_m(x),
\end{equation}
while the net flux becomes $\int_{-1}^1\sigma(s)ds$. This kernel can be expanded for small $\Peclet$ in the form
\begin{eqnarray}
    \tilde{G}=\tilde{G}_0+\Peclet \tilde{G}_1+\Peclet^2 \tilde{G}_2+\it{o}\left(\Peclet^2\right),\\
    \tilde{G}_0(x-s)=-\log{\left(\frac{\Peclet}{2}|x-s|\right)-\gamma},\\
    \tilde{G}_1(x-s)=\left(x-s\right)\left(-\log{\left(\frac{\Peclet}{2}|x-s|\right)-\gamma}\right),\\
    \tilde{G}_2(x-s)=\frac{\left(x-s\right)^2}{4}\left(C-3\log{\left(|x-s|\right)}\right),\\
    C=1-3\gamma-3\log{\left(\frac{\Peclet}{2}\right)}.
\end{eqnarray}
Note that each of $\tilde{G}_k$ is of order at most $\log{\left(\Peclet\right)}$, so that each subsequent term in our kernel expansion is indeed asymptotically smaller than its predecessor. We proceed by solving for the flux density perturbatively in $\Peclet$ where $\sigma^{\left(k\right)}(x)=\sigma_0(x)+\Peclet\sigma_1(x)+\cdots+ \Peclet^k \sigma_k$ represents the asymptotic solution to the integral equation with the approximate kernel $\tilde{G}_0+\Peclet \tilde{G}_1+\cdots+ \Peclet^k \tilde{G}_k$. In the following section, we consider the solution $\sigma^{\left(1\right)}{\left(x\right)}=\sigma_0+\Peclet \sigma_1$ when $m=1$ in which case the right side of \eqref{eq:lowpeinteq} becomes $2\pi T_1(x)$. 

\subsubsection{Leading order solution for $T_1(x)$ data}\label{leadingT1}
In the leading order problem governing $\sigma_0(x)$, the kernel simplifies to $\tilde{G}_0$ and the integral equation takes the form
\begin{eqnarray}\label{eq:T1sig0}
    \int_{-1}^1 \left(-\log{\left(\frac{\Peclet}{2}|x-s|\right)-\gamma}\right)\sigma_0(s)ds=2\pi T_1(x),
\end{eqnarray}
which can be solved exactly as follows. We begin by expanding $\sigma_0$ in a Chebyshev series,
\begin{eqnarray}
\sigma_0(x)=\frac{\sum_{k=0}^{\infty}\alpha_k^{\left(0\right)} T_k(x)}{\sqrt{1-x^2}},
\end{eqnarray}
and make note of the following important identities,
\begin{eqnarray}
\int_{-1}^1\frac{\log{\left(|x-s|\right)} T_{j}(s)}{\sqrt{1-s^2}}=-\frac{\pi}{j}T_j{\left(x\right)},\\
\int_{-1}^1\frac{\log{\left(|x-s|\right)} }{\sqrt{1-s^2}}=-\pi \log{\left(2\right)},
\end{eqnarray}
valid for $j$ being a positive integer. The solution to \eqref{eq:T1sig0} can be found by substituting our Chebyshev expansion into the integral equation, which then becomes
\begin{eqnarray}
\pi\left(\log{\left(\frac{\Peclet}{4}\right)}+\gamma\right)\alpha_0-\sum_{k=1}^{\infty}\frac{\pi}{k}\alpha_k T_k(x)=-2\pi T_1(x).\\
\end{eqnarray}
We immediately deduce the coefficients by inspection as follows,
\begin{eqnarray}
\alpha_{0}^{\left(0\right)}=0,\\
\alpha_1^{\left(0\right)}=2,\\
\alpha_{k\neq1}^{\left(0\right)}=0.
\end{eqnarray}
so that $\sigma_0$ is given explicitly  by,
\begin{equation}\label{eq:T1sig0sol}
\sigma_0(x)=\frac{2T_1(x)}{\sqrt{1-x^2}}.
\end{equation}
A key observation is that $\sigma_0$ has identically zero flux, $\int_{-1}^{1}\sigma_0{x}dx=0$, since $\alpha^{\left(0\right)}_0=0$ in its Chebyshev expansion. This means that flux can only appear at higher order in $\Peclet$.

\subsubsection{First order solution for $T_1(x)$ data}\label{FirstT1}
The solution at order $\Peclet$ is found by considering now the kernel $\tilde{G}_0+\Peclet \tilde{G}_1$. Seeking the solution to the equation of the form, $\sigma^{(1)}=\sigma_0+\Peclet \sigma_1$, the governing equation for $\sigma_1$ becomes,
\begin{equation}\label{eq:T1sig1}
    \int_{-1}^1\tilde{G}_0(x-s)\sigma_1(s)ds=-\int_{-1}^1\tilde{G}_1(x-s)\sigma_0(s)ds,
\end{equation}
where the right side of the  equation can computed using the known form of $\sigma_0$ from \eqref{eq:T1sig0sol}. We again proceed by expanding $\sigma_1$ as a Chebyshev series,
\begin{eqnarray}
\sigma_1(x)=\frac{\sum_{k=0}^{\infty}\alpha_k^{\left(1\right)} T_k(x)}{\sqrt{1-x^2}},
\end{eqnarray}
and our goal will be to deduce the coefficients $\alpha_k^{\left(1\right)}$. After performing integrals and manipulations that have been relegated to Appendix \ref{T1BCIntegrals}, we find that the integral equation reduces to the following form,
\begin{equation}
    \begin{split}
    -\pi\left(\log{\left(\frac{\Peclet}{4}\right)}+\gamma\right)\alpha_0+\sum_{k=1}^{\infty}\frac{\pi}{k}\alpha_k T_k(x)\\
    =-\frac{\pi}{2}T_2(x)-\pi\left(\log{\left(\frac{\Peclet}{4}\right)}+\gamma+1\right)T_0(x),
    \end{split}
\end{equation}
By inspection, we deduce the coefficients of the expansion for $\sigma_1$ as follows,
\begin{eqnarray}
\alpha_0^{\left(1\right)}=\frac{\left(\log{\left(\frac{\Peclet}{4}\right)}+\gamma+1\right)}{\left(\log{\left(\frac{\Peclet}{4}\right)}+\gamma\right)},\\
    \alpha_1^{\left(1\right)}=0,\\
    \alpha_2^{\left(1\right)}=-1,\\
    \alpha_{k>2}^{\left(1\right)}=0,\\
\end{eqnarray}
leading to the following expression for the flux,
\begin{equation}
\sigma^{\left(1\right)}(s)\sim \frac{2T_1(s)+\Peclet\left(\frac{\left(\log{\left(\frac{\Peclet}{4}\right)}+\gamma+1\right)}{\left(\log{\left(\frac{\Peclet}{4}\right)}+\gamma\right)}T_0(x)-T_2(x)\right)}{\sqrt{1-s^2}}.
\end{equation}
The flux is then given by the integral of $\sigma^{\left(1\right)}(x)$ over the interval $(-1,1)$, such that the flux at this order is,
\begin{equation}\label{eq:T1fluxexpression}
    F_{m=1}=\pi \Peclet \frac{\left(\log{\left(\frac{\Peclet}{4}\right)}+\gamma+1\right)}{\left(\log{\left(\frac{\Peclet}{4}\right)}+\gamma\right)}.
\end{equation}
Note that when the logarithmic correction is ignored the flux \eqref{eq:T1fluxexpression} reduces to $\pi \Peclet/2$. In practice, the logarithmic terms are significant unless $\Peclet$ is very small. We proceed by computing the flux in the case of the boundary condition $T_2(x)$. 

\subsubsection{Leading order solution for $T_2(x)$ data}\label{leadingT2}
By an analysis identical to that presented in \S \ref{leadingT1}, we find the flux density at leading order for boundary data $T_2(x)$ on the slit is given by,
\begin{equation}
\sigma_0(x)=\frac{4T_2(x)}{\sqrt{1-x^2}}.
\end{equation}
\subsubsection{First order solution for $T_2(x)$ data}
The equation governing the first order solution is then,
\begin{equation}\label{eq:T2first}
\int_{-1}^1\tilde{G}_0(x-s)\sigma_1(s)ds=-\int_{-1}^1\tilde{G}_1(x-s)\sigma_0(s)ds,
\end{equation}
which we solve by expanding $\sigma_1$ in a Chebyshev series and explicitly computing the right side of \eqref{eq:T2first}, the details of which can be found in Appendix \ref{T2BCIntegrals}. After comparing the Chebyshev coefficients in the expansion of $\sigma_1$, and the terms from the explicit computation of the right side of \eqref{eq:T2first}, we find,
\begin{equation}
    \sigma_1(x)=\frac{T_1(x)-T_3(x)}{\sqrt{1-x^2}}.
\end{equation}
Again, we observe no flux at this order of the equation. To find the flux which appears at $\it{o}\left(\Peclet\right)$, we must examine the next order kernel equation.

\subsubsection{Second order solution for $T_2(x)$ data}
The second order kernel equation takes the following form,
\begin{equation}\label{eq:T2Second}
    \int_{-1}^1\tilde{G}_0(x-s)\sigma_2(s)ds=\underbrace{-\int_{-1}^1\tilde{G}_1(x-s)\sigma_1(s)ds}_{I_1}+\underbrace{-\int_{-1}^1\tilde{G}_2(x-s)\sigma_0(s)ds}_{I_2},
\end{equation}
where the integrals $I_1$ and $I_2$ can be computed explicitly since we have already computed $\sigma_0$ and $\sigma_1$. Instead of computing both sides of \eqref{eq:T2Second} as a function of $x$, we can integrate both sides of the equation against $1/\sqrt{1-x^2}$ to obtain the flux,
\begin{equation}
-\pi\left(\log{\left(\frac{\Peclet}{4}\right)}+\gamma\right)F_{m=2}=\int_{-1}^1\frac{I_1(x)}{\sqrt{1-x^2}}dx+\int_{-1}^1\frac{I_2(x)}{\sqrt{1-x^2}}dx.
\end{equation}
Performing the integration over $x$ first greatly simplifies the integral computation once one exploits the orthogonality of the Chebyshev polynomials in the subsequent integrals over $s$. Details of the integral computations are suppressed for brevity.
Such integrals yield the following leading order expression for the total flux in the case of $m=2$,
\begin{eqnarray}\label{eq:T2FluxExpressionFinal}
    F_{m=2}\sim-\frac{\pi \Peclet^2}{4}\left(\frac{\log{\left(\frac{\Peclet}{4}\right)}+\gamma+\frac{3}{2}}{\log{\left(\frac{\Peclet}{4}\right)+\gamma}}\right),
\end{eqnarray}
in the limit of $\Peclet\rightarrow 0$.

\subsubsection{Generic boundary conditions $T_m(x)$}
Expressions for the leading order flux in the case of boundary data $T_m(x)$ with $m>2$ may be computed in a perturbative manner similar to our analysis which yielded the fluxes when $m=1$ and $m=2$. However, the perturbation equations becomes increasingly cumbersome as $m$ is increased. However, by studying and exploiting the structure of the perturbation equations, we can derive a closed form expression for the flux valid for any $m>0$.

The first important deduction is that when the boundary data is given by $T_m(x)$, the flux scales as $\Peclet^m$ asymptotically as $\Peclet\rightarrow 0$. This fact can be seen as follows. The leading order equation has solution for $m>0$ of the form,
\begin{equation}
    \sigma_0(x)=\frac{2m T_m(x)}{\sqrt{1-x^2}}.
\end{equation}
The next order equation can also be solved for arbitrary $m$ and one finds for $m>1$,
\begin{equation}
\sigma_1(x)=\frac{T_{m-1}\left(x\right)-T_{m+1}\left(x\right)}{\sqrt{1-x^2}}.
\end{equation}
In general, by examining the form of the $k^{\mathrm{th}}$-order integral equation– and noting the identity $t T_k(x)=\left(T_{k-1}\left(x\right)+T_{k-1}\left(x\right)\right)/2$, it can be seen that Chebyshev polynomials are introduced at each order in a manner such that,
\begin{equation}
\sigma_k(x)=\frac{\sum_{p=m-k}^{m+k}\alpha_p^{\left(k\right)} T_p(x)}{\sqrt{1-x^2}},
\end{equation}
where the specific values of the coefficients $\alpha_p^{\left(k\right)}$ may be deduced by performing the asymptotic analysis to order $\Peclet^k$. The key takeaway from this observation is that the flux vanishes at all orders $\Peclet^k$ for $k<m$, since the flux associated with $\sigma_k$ is simply given by $ \alpha_0^{\left(k\right)} \pi \Peclet^k $. 
Structurally, it can be seen that the $\alpha_{m}^{\left(0\right)}$ coefficient feeds the $\alpha_{m-1}^{\left(1\right)}$ coefficient in the $\sigma_k$ equation. 
Furthermore, the $\alpha_{m-1}^{\left(1\right)}$ coefficient feeds the $\alpha_{m-2}^{\left(2\right)}$, and generally the $\alpha_{m-k-1}^{\left(k-1\right)}$ coefficient feeds the $\alpha_{m-k}^{\left(k\right)}$ for $k<m$. 
By observing this fact, it is possible to derive the following recursive summation expression for $\alpha_{m-k}^{\left(k\right)}$ in terms of terms $\alpha_{m-k+j}^{\left(k-j\right)}$ with $j \geq 1$, 
\begin{eqnarray}
\alpha_{m-k}^{\left(k\right)}=-\sum_{j=1}^{k}\frac{\left(-1\right)^{j}C_j }{2^j}\frac{j! \left(m-k\right)!}{\left(m-k+j\right)!}\alpha_{m-k+j}^{\left(k-j\right)};\; 0<k<m,\\
\alpha_m^{\left(0\right)}=2m,\\
C_j=\frac{1}{ j!} {}_2F_1\left(\frac{1-j}{2},-\frac{j}{2};1;1\right)\label{eq:C},
\end{eqnarray}
where ${}_2F_1$ is the generalized hypergeometric function, the details of the derivation being relegated to Appendix \ref{FluxMDerivationAppendix}. Furthermore, by examining the kernel expansion equation at order $\Peclet^m$, it can be shown that the flux associated with $\sigma_m$ is only fed by one term in each of the lower-order flux densities, $\sigma_{k<m}$. Specifically the flux of $\sigma_m$ depends only on the terms associated with the coefficients $\alpha_{m-k}^{\left(k\right)}$, where $0\leq k < m$. It is thus possible to show that the flux associated with the boundary condition $T_m(x)$ is given by,
\begin{eqnarray}
    F_{m}=-\frac{\sum_{k=0}^{m-1} \frac{\left(-1\right)^{m-k}}{2^{m-k}}\left(C_{m-k}\left(\log{\left(\frac{\Peclet}{4}\right)}+H_{m-k}\right)+\tilde{C}_{m-k}\right)\alpha^{\left(k\right)}_{m-k}}{\log{\left(\frac{\Peclet}{4}\right)}+\gamma}\pi \Peclet^m,\label{eq:FluxM}\\
    H_{j}=\sum_{k=1}^j\frac{1}{k},\\
    \tilde{C}_{k}=-\sum_{j=0}^{\floor{\frac{k}{2}}}\frac{1}{\left(k-2j\right)!}\frac{\psi\left(j+1\right)}{\left(j!\right)^2 2^{2j}}\label{eq:Ctil},
\end{eqnarray}
where $\psi$ is the Digamma function and $H_j$ is the $j^{\mathrm{th}}$ harmonic number. Again the details of the derivation can be found in Appendix \ref{FluxMDerivationAppendix}.
We numerically evaluate the coefficients in \eqref{eq:FluxM} and tabulate the results for the cases $m\in\{1,2,3,4\}$ in \tabref{tab:coefftable}. 
For these values of $m$, we have verified the expressions against our numerical results and find precise agreement.
Specifically, we observe relative  errors between our numerically computed fluxes and the scaling predictions of $\lesssim 10\, \Peclet^2$ over the range $10^{-6} \leq \Peclet \leq 10^{-2}$ and until the predicted fluxes are below our selected quadrature precision of $10^{-12}$.

\begin{table}
  \begin{center}
\def~{\hphantom{0}}
  \begin{tabular}{l@{\hspace{30pt}}c}
      $m$  & $F_m/\left(\pi \Peclet^m\right)$  \\[1pt]
      \hline
       1   & $1+\frac{1}{\left(\gamma+\log{\left(\frac{\Peclet}{4}\right)}\right)}$\\[8pt]
       2   & $-\frac{1}{4}-\frac{3}{8\left(\gamma+\log{\left(\frac{\Peclet}{4}\right)}\right)}$\\[8pt]
       3   & $\frac{3}{48}+\frac{5}{48\left(\gamma+\log{\left(\frac{\Peclet}{4}\right)}\right)}$\\
       4   & $-\frac{5}{384}-\frac{35}{1536\left(\gamma+\log{\left(\frac{\Peclet}{4}\right)}\right)}$\\
       \vdots   & \vdots\\
       7& $\frac{11}{245760}+\frac{143}{1720320\left(\gamma+\log{\left(\frac{\Peclet}{4}\right)}\right)}$
  \end{tabular}
  \caption{Leading order expressions for the flux in the case of boundary conditions $\cos{\left(m \theta\right)}$ on the unit circle, as a function of P\'eclet number, in the asymptotic limit $\Peclet\rightarrow 0$, as computed using \eqref{eq:FluxM}.} 
  \label{tab:coefftable}.
  \end{center}
\end{table}

It is worthwhile to emphasize the following interesting aspect of the analysis which led to \eqref{eq:FluxM}. The flux associated with the $m^{\mathrm{th}}$ order perturbation equation was obtained without having to explicitly compute $\sigma_k$ at all lower orders $k<m$. This was made possible through the examination of only the structure of the perturbation problem at each order and noticing a recursive relationship between a particular coefficient at each order of the problem.

\color{black}
\subsection{Multiply-Connected Configurations}\label{examps}

\begin{figure}
\centering
    \includegraphics[width=\linewidth]{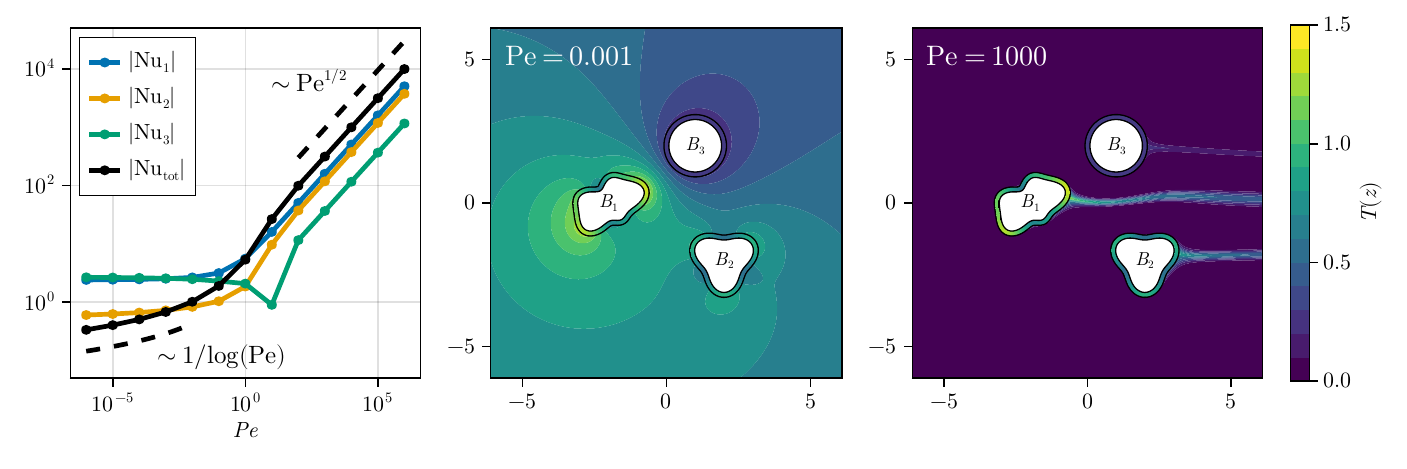}
    \caption{Left: Numerically computed heat fluxes from three obstacles with varying surface temperatures at various P\'eclet numbers.
    The dashed lines indicate the written scalings.
    In the center and right panels, we plot the temperature field for $\Peclet=10^{-3}$ and $\Peclet=10^3$.
    At low P\'eclet, the obstacles exchange heat and generally behave like a collective monopole in the far field.
    At high P\'eclet, the obstacles are thermally isolated as their wakes occupy separate streamlines.}
    \label{fig:threebody_scaling}
\end{figure}

\begin{figure}
\centering
    \includegraphics[width=\linewidth]{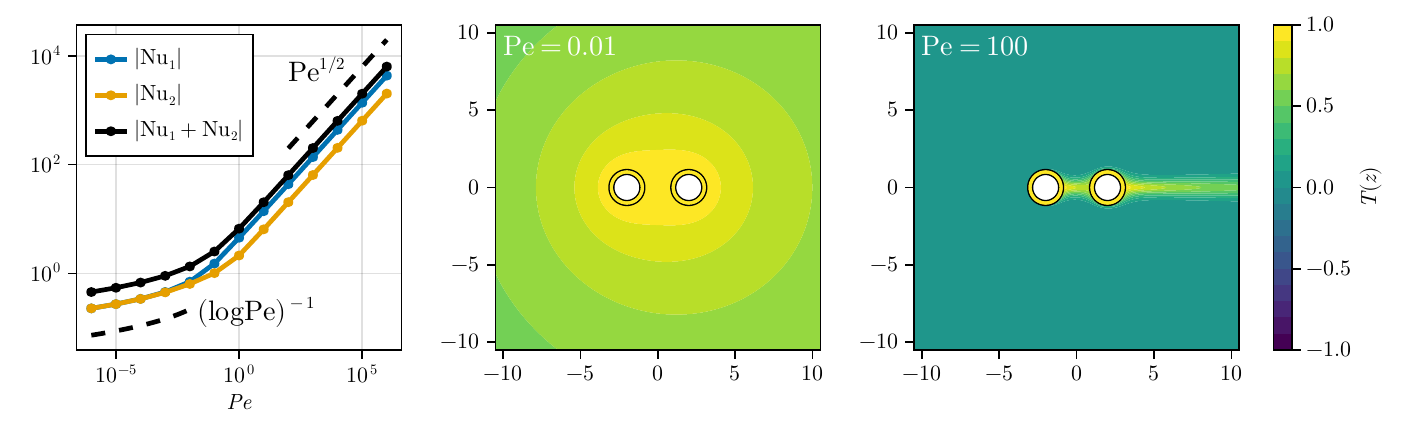}
    \includegraphics[width=\linewidth]{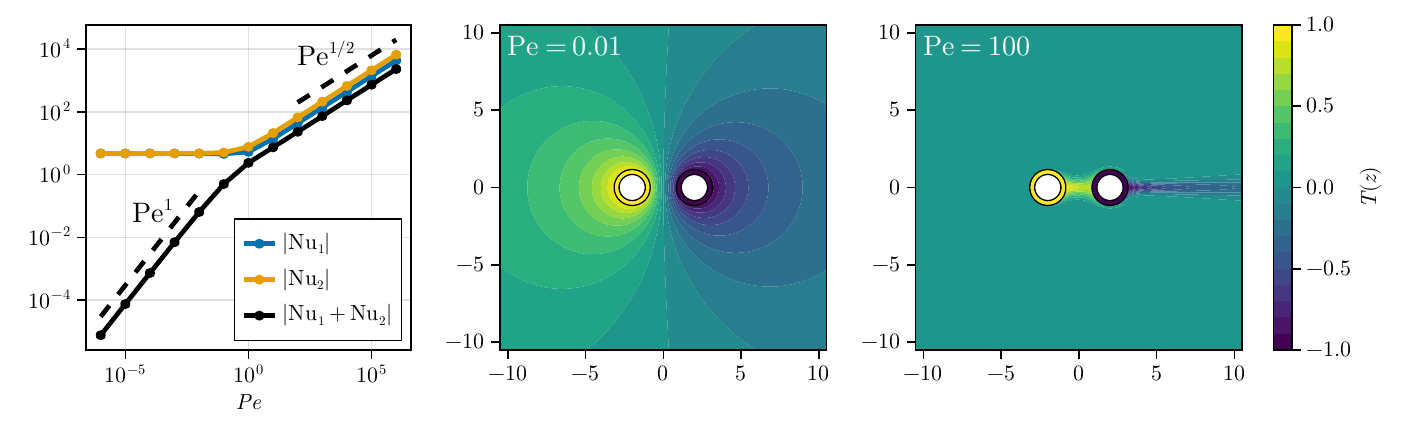}
    \caption{Left: Numerical heat fluxes from two stream-aligned disks with constant surface temperatures at various P\'eclet numbers.
    In the top row, the disks have the same temperatures, $(c_1, c_2) = (1, 1)$.
    In the bottom row, the disks have opposite temperatures, $(c_1, c_2) = (1, -1)$.
    The dashed lines indicate the written scalings.
    The center and right columns show the temperature fields for $\Peclet=10^{-2}$ and $\Peclet=10^2$.
    At low P\'eclet, the top configuration behaves like a collective monopole, while the bottom configuration approaches a Laplace solution and behaves like a collective dipole.
    At high P\'eclet, the obstacles all emit a flux following the singly-connected scaling of $\Peclet^{\frac{1}{2}}$.
    The leading order total flux at high P\'eclet can be canceled by choosing $c_2/c_1 \approx -0.467$ for this geometry.}
    \label{fig:twobody_scaling}
\end{figure}

In multiply-connected configurations, such as that presented in \figref{fig:threebodies}, obstacles can interact so as to partially cancel or enhance the heat flux out of other obstacles. 
Compared to the simply-connected case that has been studied by \cite{choi2005steady}, this produces a rich new set of inter-obstacle interactions to be studied which may have interesting implications in growth dynamics problems such as ADLA. 
Placing an obstacle downstream of another, as compared to it being placed transversely relative to the free stream direction, can particularly complicate the relative and total heat fluxes at high P\'eclet numbers.
Nonetheless, scaling regimes remain apparent in the low and high $\Peclet$ limits in multiply-connected problems.

Figure \ref{fig:threebody_scaling} depicts the fluxes at various P\'eclet numbers, as well as example temperature fields, for three generic obstacles occupying different streamlines and with varying surface temperatures.
This represents a generic case of interacting obstacles without any particular symmetries or alignment. We see that at low P\'eclet, the fluxes from each obstacle asymptotes towards constant values which cancel at leading order, and their sum asymptotes towards a multiple of the monopole ($m=0$) solution which scales like $-1/\log(\Peclet)$. For particular surface temperature distributions, for which a Laplace solution exists (with zero temperature at infinity), this leading contribution cancels for low $\Peclet$ and the flux would follow a power law scaling in P\'eclet.
At high P\'eclet, the flux from each obstacle and the total flux all scale as $\Peclet^{1/2}$, as in the simply-connected case examined by \cite{choi2005steady}. 
Physically, we can see in \figref{fig:threebody_scaling}(c), which depicts the temperature field for $\Peclet = 10^3$, that heat from each obstacle is confined to thin non-interacting wakes behind each obstacle.
In the high P\'eclet limit, the obstacles becomes thermally isolated from other obstacles on different streamlines.

In \figref{fig:twobody_scaling}, we consider two unit disks centered at $z_1=-2$ and $z_2=2$, aligned with the potential flow ($\Theta=0$).
Since the heat transfer problem is linear in the imposed obstacle temperatures, we solve the two canonical isothermal problems with $(c_1,c_2) = (1,1)$, depicted in the top row of the figure, and $(c_1,c_2)=(1,-1)$, depicted in the bottom row.
When the disks have the same temperature, they behave as a collective monopole ($m=0$) at low P\'eclet.
When the disks have opposite temperatures, there exists a Laplace solution, so the leading total flux is canceled and they behave as a collective dipole ($m=1$) at low P\'eclet.
At high P\'eclet, all of the fluxes scale as $\Peclet^\frac{1}{2}$.
However there exists a certain temperature ratio (for this geometry, $c_2/c_1 \approx -0.467$) for which the leading order fluxes will cancel.
Typically, however, even when obstacles occupy the same streamline, the total flux will follow the $\Peclet^\frac{1}{2}$ scaling.

\section{Discussion and Conclusion}

In this paper, we developed an efficient numerical procedure for solving advection-diffusion problems involving a uniform potential flow past a collection of impermeable obstacles with generic boundary temperature distributions. 
Our work thus extends prior studies, which examined advection-diffusion in flows past a single obstacle with isothermal boundary conditions. 
Our method leverages the AAA-LS method for solving Laplace problems \citep{costa2021aaa}, to efficiently compute the complex potential of the corresponding potential flow, which doubles as a conformal mapping to our computation domain. 
In the mapped domain, comprising a collection of horizontal slits in an otherwise infinite plane, we formulate a mixed boundary integral solution to the heat transfer problem with arbitrary Dirichlet data on each boundary. 
We thus enable an efficient solution to advection-diffusion in potential flows past arbitrary collections of obstacles. 
We note that our numerical approach is fast and highly accurate when obstacles are smooth and not too close in proximity to one another. 

We then applied this method to examine the heat flux from obstacles in various heat-transfer regimes.
In the limit of high P\'eclet numbers, we demonstrate in \ref{fig:lowpecosinesplot}(a) that the flux exiting each obstacle scales as $\Peclet^{1/2}$, in agreement with the scaling for a simply-connected isothermal obstacle given in (\ref{eq:choloePeNuss}). 
A notable exception occurs in the simply-connected problem for boundary conditions of the type $\sin(m\eta)$ in the Joukowsky parametrization, in which case $\mathcal{F}=0$ independent of P\'eclet number.

We showed numerically that the Nusselt number at low P\'eclet numbers, with variable boundary data, no longer scales in the same manner as the simply-connected isothermal problem (\ref{eq:choloePeNuss}). 
We then derived asymptotic Nusselt number scalings in the limit of low P\'eclet numbers for flow past a circle with arbitrary boundary data. 
In particular, we show that boundary data of the form $\cos\left(m\theta\right)$ on the unit circle produces a net flux in proportion to $\Peclet^m$ for $m \geq 1$. Expressions for the leading coefficient of the scaling were derived for all cases $m\geq1$ using a perturbative approach leading to \eqref{eq:FluxM}, with explicit expressions for the cases $m\in\{1,2,3,4\}$ given in \tabref{tab:coefftable}. These expressions include logarithmic corrections and are found to agree to high precision with our numerical results, specifically with relative errors $\approx \Peclet^2$ for $\Peclet \leq 10^{-2}$ and above our set precision limit ($|\Nusselt| > 10^{-12}$).
We have thus shown that, in the limit of small $\Peclet$, the flux out of an arbitrary obstacle is determined by its Chebyshev representation in the corresponding slit domain, or equivalently its cosine series under the Joukowsky parametrization.

Our results might prove useful in a variety of future studies of advection-diffusion within potential flows. 
Because we employ a boundary integral approach, our solution procedure is highly accurate and is substantially faster than volume-based methods. 
In the future, we hope to extend our results to more general types of potential flows, which might include circulation and flow singularities in addition to a free stream. 
Such extensions would make possible investigations of (\ref{eq:st_addiff}) within realistic flows such as the lifting flow around a wing and other aerodynamic applications.
Other interesting directions might include the investigation of growth, melting, and artificial freezing problems, which have mainly been analyzed in the context of simply-connected domains with isothermal (constant Dirichlet) boundary conditions. 
It would be interesting to see how fluid dynamically mediated interactions between obstacles might influence their growth.

Another potential application is in Hele-Shaw flows and microfluidic systems, where a common objective is to promote mixing. 
For example, one might study the evolution of solutes within Hele-Shaw flows; one might be interested in the concentration of important biological species in lab-on-a-chip technologies or possibly the evolution of a simple dye. 
Our method may also be useful in the design of cold plate systems meant to optimize heat transfer, in systems where the approximation of potential flow is reasonable, as our method obviates the need for solving the full conjugate heat transfer equations or resorting to surrogate models during optimization.
Again, extensions of the present work to generalized potential flows -- i.e., flows with circulation or point sources/sinks/vortices -- would be advantageous. 
While some work has investigated steady advection diffusion in point source/sink flows, the effects of obstacles has not yet to be studied \citep{boulais2020two}. 
Magnetohydrodynamic Hele-Shaw flows \citep{mckee2024magnetohydrodynamic} represent singularity-free potential flows which in general comprise both a free stream as well as circulation around obstacles. 
Since one possible application of such flows is in promoting mixing, monitoring advection diffusion processes is certainly of interest.
Thus, we intend to extend our present framework to treat such flows in the future.

\paragraph{\textbf{Acknowledgements.} The authors are grateful for many valuable discussions with Nick Trefethen, Daniel Fortunato, Alex Barnett, Leslie Greengard, Sheehan Olver, Ruben Rosales, Toby Driscoll, Darren Crowdy, Hai Zhu, and Martin Bazant.}\\
\\
\paragraph{\textbf{Funding.} K.M was generously funded by a Mathworks fellowship during this work. This work was supported by a grant from the Simons Foundation (SSRFA-6826, K.J.B.).}\\
\\
\paragraph{\textbf{Declaration of Interests.} } The authors report no conflict of interest.

\appendix
\section{Integrals from $T_1(x)$ Perturbation Analysis}\label{T1BCIntegrals}
The right side of \eqref{eq:T1sig1} which we define by $I_R=\left(-\int_{-1}^1\tilde{G}_1(x-s)\sigma_0(s)ds\right)$ is written explicitly as follows,
\begin{equation}
    I_R =2\int_{-1}^1  \left(x-s\right)\left(\log{\left(|x-s|\right)}+\log{\left(\frac{\Peclet}{2}\right)+\gamma}\right) \frac{T_1(s)}{\sqrt{1-s^2}}ds.
\end{equation}
The first term of the expression becomes,
\begin{equation}
\begin{split}
    2\int_{-1}^1  \frac{\left(x-s\right)T_1(s)}{\sqrt{1-s^2}}\log{\left(|x-s|\right)}ds=-2\pi x T_1(x)-2\int_{-1}^1\frac{T_1(s)T_1(s)\log{\left(|x-s|\right)}}{\sqrt{1-s^2}}\\
    =-\pi\left(T_0\left(x\right)+T_2\left(x\right)\right)+\pi\left( \log{\left(2\right)}T_0(x)+\frac{1}{2}T_{2}\left(x\right)\right)\\
    =\pi\left(\log{\left(2\right)}-1\right)-\frac{\pi}{2}T_2(x),
\end{split}
\end{equation}
and the second term of which becomes,
\begin{eqnarray}
    2\left(\log{\left(\frac{\Peclet}{2}\right)}+\gamma\right)\int_{-1}^1  \frac{\left(x-s\right)T_1(s)}{\sqrt{1-s^2}}ds=-\pi\left(\log{\left(\frac{\Peclet}{2}\right)}+\gamma\right)T_0(x),
\end{eqnarray}
so that in sum
\begin{eqnarray}
    I_R=-\frac{\pi}{2} T_2(x)-\pi\left(\log{\left(\frac{\Peclet}{4}\right)}+\gamma+1\right)T_0(x).
\end{eqnarray}

\section{Integrals from $T_2(x)$ Perturbation Analysis}\label{T2BCIntegrals}
\subsection{Integrals appearing in first order equation}
After noting that all $s$-independent terms of the kernel vanish due to the orthogonality of Chebyshev polynomials, the right term of \eqref{eq:T2first} becomes,
\begin{equation}
\begin{split}
4\int_{-1}^1\left(x-s\right)\frac{T_2(s)\log{\left(|x-s|\right)}}{\sqrt{1-s^2}}ds=-4x\frac{\pi}{2}T_2(x)+2\pi\left(\frac{1}{3}T_3(x)+T_1(x)\right)\\
=-\pi\left(T_1(x)+T_3(x)\right)+2\pi\left(\frac{1}{3}T_3(x)+T_1(x)\right)\\
=\pi\left(-\frac{1}{3}T_3(x)+T_1(x)\right).
\end{split}
\end{equation}

\section{Derivation of flux equation for $T_m(x)$ data}\label{FluxMDerivationAppendix}
\subsection{Determination of expression for $\alpha_{m-k}^{\left(k\right)}$ for $k<m$}
We begin by examining the $k^{\mathrm{th}}$ order integral equation which governs the flux density $\sigma_k$ which takes the form,
\begin{equation}\label{eq:sigkequation}
    \int_{-1}^1\tilde{G}_0(x-s)\sigma_k(s)ds=-\sum_{p=1}^k\int_{-1}^1\tilde{G}_p(x-s)\sigma_{k-p}(s)ds,
\end{equation}
where we note that $\sigma_p$ for $p<k$ are known, at this order of perturbation theory, via the solution of lower order perturbation equations. We proceed by considering the Chebyshev expansions of $\sigma_j$ for all $j$ and isolating the terms associated with $T_{m-k}$ on both sides of the equation so that $\alpha_{m-k}^{\left(k\right)}$ may be deduced. To this end, for the remainder of this section of the paper we shall suppress all terms that do not contribute to terms of the form $T_{m-k}$.

Let us isolate each term in the right side of the equation by defining,
\begin{equation}
    I_p\equiv -\int_{-1}^1 \tilde{G}_p(x-s)\sigma_{p-s}ds,
\end{equation}
so the right side of \eqref{eq:sigkequation} is equal to the sum of all the $I_p$ integrals over the range $1\leq p \leq k$. Writing out the kernel explicity gives the following expression for the portion of $I_p$ that can contribute terms proportional to $T_{m-k}\left(x\right)$,
\begin{equation}\label{eq:IpExpanded}
    I_p = \int_{-1}^1\left(x-s\right)^p\left(C_p \log{\left(\frac{\Peclet}{2}|x-s|\right)}+\tilde{C}_p\right)\frac{\alpha_{m-k+p}^{\left(k-p\right)} T_{m-k+p}\left(s\right)}{\sqrt{1-s^2}},
\end{equation}
where $C_p$ and $\tilde{C}_p$ represents the $p^{\mathrm{th}}$ coefficients of the following expansions,
\begin{eqnarray}
e^{\Peclet x}I_0{\left(\Peclet x\right)}=\sum_{p=0}^{\infty}C_p\Peclet^p x^p,\\
-e^{\Peclet x} \sum_{j=0}^{\infty}\psi\left(j+1\right)/\left(j!\right)^2 \left(\Peclet x/2\right)^{2j}=\sum_{p=0}^{\infty}\tilde{C}_p\Peclet^p x^p
\end{eqnarray}
which are valid for small $\Peclet$. The forms of $C_p$ and $\tilde{C}_p$ are also summarized in the main text in \eqref{eq:C} and \eqref{eq:Ctil}. Once again, we note that \eqref{eq:IpExpanded} only contains the contribution to the integral that is relevant in producing terms of the type $T_{m-k}\left(x\right)$.

Now since $m-k\geq1$, all terms in the integral not associated with the $\log\left(|x-s|\right)$ term must vanish by the orthogonality of Chebyshev polynomials of unequal indices. It follows that $I_p$ reduces to the following form,
\begin{equation}\label{eq:IpReduced}
    I_p = C_p \alpha_{m-k+p}^{\left(k-p\right)} \int_{-1}^1 \frac{ \left(x-s\right)^p T_{m-k+p}\left(s\right)\log{\left(|x-s|\right)}}{\sqrt{1-s^2}}.
\end{equation}
Every term in the expansion of $\left(x-s\right)^p$ is capable of generating terms of the form $T_{m-k}(x)$ through integration. By expanding $\left(x-s\right)^p$ and employing integral identities of the Chebyshev polynomials, one can show that,
\begin{eqnarray}
    I_p=C_p \alpha_{m-k+p}^{\left(k-p\right)} \sum_{l=0}^{p}\frac{\left(-1\right)^l p! }{l! \left(p-l\right)!}x^{p-l}\int_{-1}^1\frac{ s^l T_{m-k+p}\left(s\right)\log{\left(|x-s|\right)}}{\sqrt{1-s^2}}\\
    =C_p \alpha_{m-k+p}^{\left(k-p\right)} \sum_{l=0}^{p}\frac{\left(-1\right)^l p! }{l! \left(p-l\right)!}x^{p-l}\frac{-\pi}{2^l \left(m-k+p-l\right)}T_{m-k+p-l}\left(x\right)\\
    =-\frac{\pi C_p T_{m-k}\left(x\right) \alpha_{m-k+p}^{\left(k-p\right)}}{2^p} \sum_{l=0}^{p}\frac{\left(-1\right)^l p! }{l! \left(p-l\right)!}\frac{1}{ \left(m-k+p-l\right)},
\end{eqnarray}
where the last summation can be evaluated explicitly and is given as follows, $\left(-1\right)^p p!\left(m-k-1\right)!/\left(m-k+p\right)!$. Noting that the relevant term in the left side of \eqref{eq:sigkequation} reduces to $\frac{\pi}{m-k} \alpha_{m-k}^{\left(k\right)} T_{m-k}\left(x\right)$, we find the following expression for $\alpha_{m-k}^{\left(k\right)}$,
\begin{eqnarray}
\alpha_{m-k}^{\left(k\right)}=-\sum_{p=1}^{k}\frac{ C_p  \alpha_{m-k+p}^{\left(k-p\right)}}{2^p}\frac{\left(-1\right)^p p! \left(m-k\right)!}{\left(m-k+p\right)!}\alpha_{m-k+p}^{\left(k-p\right)};\;0<k<m,\\
\alpha_{m}^{\left(0\right)}=2m,
\end{eqnarray}
the latter equation setting the base case of the recursive relation for the coefficients $\alpha_{m-k}^{\left(k\right)}$ with $k>0$. This expression can be evaluated symbolically using just a few lines of code in standard software such as Mathematica.

\subsection{Determination of flux at order $\Peclet^m$}
The $\it{O}\left(\Peclet^m \right)$ equations read as follows,
\begin{equation}\label{eq:sigmequation}
    \int_{-1}^1\tilde{G}_0(x-s)\sigma_m(s)ds=\sum_{k=0}^{m-1}\int_{-1}^1\underbrace{-\tilde{G}_{m-k}(x-s)\sigma_{k}(s)ds}_{I_k},
\end{equation}
where the portion of $I_k$ that may contribute to the flux takes the form,
\begin{equation}
    I_k = \int_{-1}^1\left(C_{m-k} \log{\left(\frac{\Peclet}{2}|x-s|\right)}+\tilde{C}_{m-k}\right)\frac{\alpha_{m-k}^{\left(k\right)} T_{m-k}\left(s\right)\left(x-s\right)^{m-k}}{\sqrt{1-s^2}},
\end{equation}
which can be decomposed as follows,
\begin{eqnarray}
    I_k=C_{m-k}\alpha_{m-k}^{\left(k\right)} I_k'+\left(\tilde{C}_{m-k}+C_{m-k}\log{\left(\frac{\Peclet}{2}\right)}\right) \alpha_{m-k}^{\left(k\right)} I_k'',\\
    I_k' = \int_{-1}^1\log{\left(|x-s|\right)}\frac{ T_{m-k}\left(s\right)\left(x-s\right)^{m-k}}{\sqrt{1-s^2}},\\
    I_k'' = \int_{-1}^1\frac{ T_{m-k}\left(s\right)\left(x-s\right)^{m-k}}{\sqrt{1-s^2}}.
\end{eqnarray}
By the orthogonality of Chebyshev polynomials, $I_2''$ vanishes for every term in the expansion of $\left(x-s\right)^{m-k}$ except the $s^{m-k}$ term such that,
\begin{equation}
I_k''=\frac{\pi}{2^{m-k}}\left(-1\right)^{m-k},
\end{equation}
whereas all the terms in $I_k'$ will need to be summed. Writing out the expansion of $I_k'$ explicitly, we find that 
\begin{equation}
I_k'=\sum_{p=0}^{m-k}\frac{\left(-1\right)^{p}\left(m-k\right)! x^{m-k-p}}{p! \left(m-k-p\right)!}\int_{-1}^1\log{\left(|x-s|\right)}\frac{ T_{m-k}\left(s\right)s^p}{\sqrt{1-s^2}},
\end{equation}
which reduces to
\begin{eqnarray}
I_k'=-\frac{\pi}{2^{m-k}} \sum_{p=0}^{m-k-1}\frac{\left(-1\right)^{p}\left(m-k\right)!}{p! \left(m-k-p\right)!}\frac{1}{m-k-p}-\frac{\left(-1\right)^{m-k}\pi \log{2}}{2^{m-k}},
\end{eqnarray}
where the first sum can be computed exactly as $\left(-1\right)^{m-k+1} H_{m-k}$, where $H_k$ is the $k^{\mathrm{th}}$ harmonic number. We thus find that,
\begin{equation}
    I_k'=\frac{\pi \left(-1\right)^{m-k}H_{m-k}-\left(-1\right)^{m-k}\pi \log{2}}{2^{m-k}}.
\end{equation}
Summing $I_k'$ and $I_k''$, we find that the coefficient of the constant term associated with the right side of \eqref{eq:sigmequation} is given by,
\begin{equation}
    \sum_{k=0}^{m-1} \frac{\pi \left(-1\right)^{m-k}}{2^{m-k}}\left(C_{m-k}\left(\log{\left(\frac{\Peclet}{4}\right)}+H_{m-k}\right)+\tilde{C}_{m-k}\right)\alpha_{m-k}^{\left(k\right)}.
\end{equation}
After comparing this constant with the left side of \eqref{eq:sigmequation}, we find that 
\begin{equation}
    \alpha_{0}^{\left(m\right)}=-\frac{\sum_{k=0}^{m-1}\frac{ \left(-1\right)^{m-k}}{2^{m-k}}\left(C_{m-k}\left(\log{\left(\frac{\Peclet}{4}\right)}+H_{m-k}\right)+\tilde{C}_{m-k}\right)\alpha_{m-k}^{\left(k\right)}}{\log{\left(\frac{\Peclet}{4}\right)}+\gamma}.
\end{equation}
The flux is then simply given by $\pi \alpha_{0}^{\left(m\right)} \Peclet^m$.

\bibliographystyle{jfm}
\bibliography{manuscript_text}

\begin{thebibliography}{28}
\expandafter\ifx\csname natexlab\endcsname\relax\def\natexlab#1{#1}\fi
\def\au#1{#1} \def\ed#1{#1} \def\yr#1{#1}\def\at#1{#1}\def\jt#1{\textit{#1}} \def\bt#1{#1}\def\bvol#1{\textbf{#1}} \def\vol#1{#1} \def\pg#1{#1} \def\publ#1{#1}\def\arxiv#1{#1}\def\org#1{#1}\def\st#1{\textit{#1}}

\bibitem[Alimov {\em et~al.\/}(1994)Alimov, Kornev \& Mukhamadullina]{alimov1994equilibrium}
{\sc \au{Alimov, MM}, \au{Kornev, KG} \& \au{Mukhamadullina, GL}} \yr{1994}  \at{The equilibrium shape of an ice-soil body formed by liquid flow past a pair of freezing columns}.  \jt{Journal of Applied Mathematics and Mechanics}  \bvol{58}~(5),  \pg{873--888}.

\bibitem[Atkinson \& Sloan(1991)]{atkinson1991}
{\sc \au{Atkinson, Kendall~E} \& \au{Sloan, Ian~H}} \yr{1991}  \at{The numerical solution of first-kind logarithmic-kernel integral equations on smooth open arcs}.  \jt{Mathematics of computation}  \bvol{56}~(193),  \pg{119--139}.

\bibitem[Baddoo(2020)]{baddoo2020lightning}
{\sc \au{Baddoo, Peter~J}} \yr{2020}  \at{Lightning solvers for potential flows}.  \jt{Fluids}  \bvol{5}~(4),  \pg{227}.

\bibitem[Bazant(2004)]{bazant2004conformal}
{\sc \au{Bazant, Martin~Z}} \yr{2004}  \at{Conformal mapping of some non-harmonic functions in transport theory}.  \jt{Proceedings of the Royal Society of London. Series A: Mathematical, Physical and Engineering Sciences}  \bvol{460}~(2045),  \pg{1433--1452}.

\bibitem[Bazant {\em et~al.\/}(2003)Bazant, Choi \& Davidovitch]{bazant2003dynamics}
{\sc \au{Bazant, Martin~Z}, \au{Choi, Jaehyuk} \& \au{Davidovitch, Benny}} \yr{2003}  \at{Dynamics of conformal maps for a class of non-laplacian growth phenomena}.  \jt{Physical review letters}  \bvol{91}~(4),  \pg{045503}.

\bibitem[Bazant \& Moffatt(2005)]{bazant2005exact}
{\sc \au{Bazant, Martin~Z} \& \au{Moffatt, HK}} \yr{2005}  \at{Exact solutions of the navier--stokes equations having steady vortex structures}.  \jt{Journal of Fluid Mechanics}  \bvol{541},  \pg{55--64}.

\bibitem[Boulais \& Gervais(2020)]{boulais2020two}
{\sc \au{Boulais, Etienne} \& \au{Gervais, Thomas}} \yr{2020}  \at{Two-dimensional convection--diffusion in multipolar flows with applications in microfluidics and groundwater flow}.  \jt{Physics of Fluids}  \bvol{32}~(12).

\bibitem[Boussinesq(1902)]{boussinesq1902pouvoir}
{\sc \au{Boussinesq, J}} \yr{1902}  \at{Sur le pouvoir refroidissant d'un courant liquide ou gazeux}.  \jt{J. Phys. Theor. Appl.}  \bvol{1}~(1),  \pg{71--75}.

\bibitem[Boussinesq(1905)]{boussinesq1905calculation}
{\sc \au{Boussinesq, J}} \yr{1905}  \at{Calculation of the cooling power of fluid currents}.  \jt{Journal of pure and applied mathematics}  \bvol{1},  \pg{285--332}.

\bibitem[Brubeck {\em et~al.\/}(2021)Brubeck, Nakatsukasa \& Trefethen]{brubeck2021vandermonde}
{\sc \au{Brubeck, Pablo~D}, \au{Nakatsukasa, Yuji} \& \au{Trefethen, Lloyd~N}} \yr{2021}  \at{Vandermonde with arnoldi}.  \jt{Siam Review}  \bvol{63}~(2),  \pg{405--415}.

\bibitem[Cheng \& Greengard(1998)]{cheng_method_1998}
{\sc \au{Cheng, Hongwei} \& \au{Greengard, Leslie}} \yr{1998}  \at{A {Method} of {Images} for the {Evaluation} of {Electrostatic} {Fields} in {Systems} of {Closely} {Spaced} {Conducting} {Cylinders}}.  \jt{SIAM Journal on Applied Mathematics}  \bvol{58}~(1),  \pg{122--141}.

\bibitem[Choi {\em et~al.\/}(2005)Choi, Margetis, Squires \& Bazant]{choi2005steady}
{\sc \au{Choi, Jaehyuk}, \au{Margetis, Dionisios}, \au{Squires, Todd~M} \& \au{Bazant, Martin~Z}} \yr{2005}  \at{Steady advection--diffusion around finite absorbers in two-dimensional potential flows}.  \jt{Journal of Fluid Mechanics}  \bvol{536},  \pg{155--184}.

\bibitem[Costa \& Trefethen(2023)]{costa2021aaa}
{\sc \au{Costa, Stefano} \& \au{Trefethen, Lloyd~N}} \yr{2023}  \at{{AAA}-least squares rational approximation and solution of laplace problems}.  \bt{In {\em European Congress of Mathematics Portorož, 20–26 June 2021\/}},  \pg{pp. 511--534}.  \publ{Berlin, Germany: EMS Press}.

\bibitem[Crowdy(2020)]{crowdy2020solving}
{\sc \au{Crowdy, D.}} \yr{2020} {\em Solving problems in multiply connected domains\/}.  \publ{SIAM}.

\bibitem[Davidovitch {\em et~al.\/}(2005)Davidovitch, Choi \& Bazant]{davidovitch2005average}
{\sc \au{Davidovitch, Benny}, \au{Choi, Jaehyuk} \& \au{Bazant, Martin~Z}} \yr{2005}  \at{Average shape of transport-limited aggregates}.  \jt{Physical review letters}  \bvol{95}~(7),  \pg{075504}.

\bibitem[Galante \& Churchill(1990)]{galante1990applicability}
{\sc \au{Galante, Stephen~R} \& \au{Churchill, Stuart~W}} \yr{1990}  \at{Applicability of solutions for convection in potential flow}.  \bt{In {\em Advances in heat transfer\/}}, ,  \vol{vol.~20},  \pg{pp. 353--388}.  \publ{Elsevier}.

\bibitem[Gopal \& Trefethen(2019)]{gopal2019solving}
{\sc \au{Gopal, Abinand} \& \au{Trefethen, Lloyd~N}} \yr{2019}  \at{Solving laplace problems with corner singularities via rational functions}.  \jt{SIAM Journal on Numerical Analysis}  \bvol{57}~(5),  \pg{2074--2094}.

\bibitem[Goyette {\em et~al.\/}(2019)Goyette, Boulais, Normandeau, Laberge, Juncker \& Gervais]{goyette2019microfluidic}
{\sc \au{Goyette, Pierre-Alexandre}, \au{Boulais, {\'E}tienne}, \au{Normandeau, Fr{\'e}d{\'e}ric}, \au{Laberge, Gabriel}, \au{Juncker, David} \& \au{Gervais, Thomas}} \yr{2019}  \at{Microfluidic multipoles theory and applications}.  \jt{Nature communications}  \bvol{10}~(1),  \pg{1781}.

\bibitem[Hsu(1965)]{hsu1965heat}
{\sc \au{Hsu, Chia-Jung}} \yr{1965}  \at{Heat transfer to liquid metals flowing past spheres and elliptical-rod bundles}.  \jt{International Journal of Heat and Mass Transfer}  \bvol{8}~(2),  \pg{303--315}.

\bibitem[McKee(2024)]{mckee2024magnetohydrodynamic}
{\sc \au{McKee, Kyle~I}} \yr{2024}  \at{Magnetohydrodynamic flow control in {H}ele-{S}haw cells}.  \jt{Journal of Fluid Mechanics}  \bvol{993},  \pg{A11}.

\bibitem[Mukhamadullina {\em et~al.\/}(1998)Mukhamadullina, Kornev \& Alimov]{mukhamadullina1998hysteretic}
{\sc \au{Mukhamadullina, G}, \au{Kornev, K} \& \au{Alimov, M}} \yr{1998}  \at{Hysteretic effects in the problems of artificial freezing}.  \jt{SIAM Journal on Applied Mathematics}  \bvol{59}~(2),  \pg{387--410}.

\bibitem[Nakatsukasa {\em et~al.\/}(2018)Nakatsukasa, S{\`e}te \& Trefethen]{nakatsukasa2018aaa}
{\sc \au{Nakatsukasa, Yuji}, \au{S{\`e}te, Olivier} \& \au{Trefethen, Lloyd~N}} \yr{2018}  \at{The aaa algorithm for rational approximation}.  \jt{SIAM Journal on Scientific Computing}  \bvol{40}~(3),  \pg{A1494--A1522}.

\bibitem[Nehari(2012)]{nehari2012conformal}
{\sc \au{Nehari, Zeev}} \yr{2012} {\em Conformal mapping\/}.  \publ{Courier Corporation}.

\bibitem[Newman(1964)]{newman1964rational}
{\sc \au{Newman, Donald~J}} \yr{1964}  \at{Rational approximation to $| x| $.}  \jt{Michigan Mathematical Journal}  \bvol{11}~(1),  \pg{11}.

\bibitem[Rycroft \& Bazant(2016)]{rycroft2016asymmetric}
{\sc \au{Rycroft, Chris~H} \& \au{Bazant, Martin~Z}} \yr{2016}  \at{Asymmetric collapse by dissolution or melting in a uniform flow}.  \jt{Proceedings of the Royal Society A: Mathematical, Physical and Engineering Sciences}  \bvol{472}~(2185),  \pg{20150531}.

\bibitem[Trefethen(2020)]{trefethen2020numerical}
{\sc \au{Trefethen, Lloyd~N}} \yr{2020}  \at{Numerical conformal mapping with rational functions}.  \jt{Computational Methods and Function Theory}  \bvol{20},  \pg{369--387}.

\bibitem[Trefethen(2024)]{trefethen2024polynomial}
{\sc \au{Trefethen, Lloyd~N}} \yr{2024}  \at{Polynomial and rational convergence rates for laplace problems on planar domains}.  \jt{Proceedings of the Royal Society A}  \bvol{480}~(2295),  \pg{20240178}.

\bibitem[Trefethen(2025)]{trefethen2025numerical}
{\sc \au{Trefethen, Lloyd~N}} \yr{2025}  \at{Numerical computation of the schwarz function}.  \jt{arXiv preprint arXiv:2501.00898} .

\end{thebibliography}

\end{document}